\documentclass[manuscript]{aastex}

\usepackage{amsmath}
\usepackage{rotating}
\usepackage{hyperref}

\slugcomment{Accepted PASP}

\shortauthors{Iglesias-Marzoa et al.}

\begin{document}


\title{The {\it rvfit} Code: A Detailed Adaptive Simulated Annealing\\
    Code for Fitting Binaries and Exoplanets Radial Velocities}


\author{Ram\'on Iglesias-Marzoa}
\affil{Astrophysics Department, Universidad de La Laguna, E-38205 La Laguna, Tenerife, Spain}
\affil{Centro de Estudios de F\'isica del Cosmos de Arag\'on, Plaza San Juan 1, E-44001, Teruel, Spain}

\email{riglesias@cefca.es}

\and

\author{Mercedes L\'opez-Morales}
\affil{Harvard-Smithsonian Center for Astrophysics, 60 Garden Street, Cambridge, MA 02138, USA}

\and

\author{Mar\'ia Jes\'us Ar\'evalo Morales}
\affil{Astrophysics Department, Universidad de La Laguna, E-38205 La Laguna, Tenerife, Spain}
\affil{Instituto de Astrof\'isica de Canarias, E-38200 La Laguna, Tenerife, Spain}


\begin{abstract}
The fitting of radial velocity curves is a frequent procedure in binary stars
and exoplanet research. In the majority of cases the fitting routines need to be
fed with a set of initial parameter values and priors from which to begin the
computations and their results can be affected by local minima. We present a
new code, the \verb;rvfit; code, for fitting radial velocities of stellar binaries and
exoplanets using an Adaptive Simulated Annealing (ASA) global minimization
method, which fastly converges to a global solution minimum without the need
to provide preliminary parameter values. We show the performance of the code
using both synthetic and real data sets: double-lined binaries, single-lined binaries,
and exoplanet systems. In all examples the keplerian orbital parameters
fitted by the \verb;rvfit; code and their computed uncertainties are compared with
literature solutions. Finally, we provide the source code with a working example
and a detailed description on how to use it.
\end{abstract}


\section{Introduction}

Precise, absolute masses of stars and planets can be only measured if they are part of binary
(star-star or star-planet) systems. The masses are derived by fitting the radial velocity (RV) curves of
those systems to the keplerian orbital equations. Those equations need to be solved numerically using
multi-parameter minimization techniques.

The fitting of radial velocity curves is an a-priori straightforward procedure that gets complicated
by the need to explore a wide multi-parameter space and by the existence of many potential local minima, 
which can yield to incorrect solutions. Local minima are a well known issue in mathematical optimization
problems and a lot of work has been done over the years to overcome this limitation when trying to
fit multi-parameter functions. Simultaneous, multi-parameter minimization techniques, such as
the Levenberg-Marquardt (LM) algorithm \citep{levenberg1944, marquardt1963, press1992},
also known as the damped least-squares (DLS) method,
or the Nelder \& Mead Simplex (NMS) algorithm \citep{neldermead1965, murty1983},
are widely used in multi-parameter function optimization problems in many science fields.

Different minimization methods are implemented in the most widely used binary and exoplanet
modelling softwares. Some of those softwares are the Wilson-Devinney code \citep{wilson1971}
which is the {\it de facto} standard code to analyze eclipsing binary stars and implements
the Differential Corrections (DC) method which is derivative based and is used to look for
local convergence. A wrapper for the Wilson-Devinney Method, WD2007, was built by
J. Kallrath \citep{milone2008} and implements Simulated Annealing (SA)
in the Boltzmann version \citep[see e.~g][]{metropolis1953,kirkpatrick1983}.
PHOEBE \citep[PHysics Of Eclipsing BinariEs;][]{prsa2005} is a code which models light and RV
curves. PHOEBE relies in the Wilson-Devinney code and acts as a front-end interface
adding new capabilities. Another commonly used software is Nightfall \citep{wichmann2011},
which is a code to build synthetic light and RV curves of eclipsing
binary stars and can fit a model to photometric and RV data. Nightfall uses
the 'Simplex Algorithm' for local optimizacion and SA for global optimization
\citep{wichmann1999}.

In recent years, the discovery of hundreds of exoplanets using RV techniques has lead to the
development of several software packages to fit RV and transit photometric curves. One
example is the Systemic Console \citep{meschiari2009, meschiari2010}, which focuses
in the fitting of exoplanet RV and transit curves and includes the LM, NMS and SA algorithms.


All those software packages include methods for local and global minimisation, SA among them.
SA is a multi-parameter minimization technique that draws on metallurgic optimal cooling
processing methods \citep[see][]{kirkpatrick1983}, and it has been implemented and tested in some
of those popular codes. However, the technique has been somewhat demoted on claims of
being notoriously slow and less efficient than other methods
\citep[e.~g. ][]{prsa2005,kallrath2006,wichmann1999}.

Faster methods based on derivatives (LM, DC), start from points
near the solution but this assumes a previous knowledge of the solution, which is only tuned by
the algorithm.

In this paper we introduce a modified version of the SA method called Adaptive Simulated
Annealing (ASA), which overcomes the speed problem of standard SA methods and ensures fast converging
to a global minimum solution for stellar binaries and exoplanet radial velocity curves. We have
developed an ASA minimization code in IDL\footnote{IDL is a high-level commercial programming
language and environment by Exelis Visual Information Solutions.
\url{http://www.exelisvis.com/ProductsServices/IDL.aspx}},
which we make available to the community, and which
can be easily implemented as part of any custom binary or exoplanet analysis software.  We also
provide a detailed guide on how to implement this ASA method for fitting radial velocity
curves to a set of radial velocity measurements.

We describe the ASA method in section \ref{sec:ASAtheory}. In section \ref{sec:objectivefunc}
we describe the function that the algorithm aims at minimising, and the algorithm itself
is described in section \ref{sec:ASAalgorithm} and \ref{sec:uncertainties}.
We present some examples of the performance of the algorithm in section \ref{sec:test_objects},
and discuss the results and pros and cons of the algorithm in sections \ref{sec:discussion}
and \ref{sec:conclusions}. The code with all its documentation is publicly available at
\url{http://www.cefca.es/people/~riglesias/index.html}.

\section{(Adaptive) Simulated Annealing}
\label{sec:ASAtheory}

SA is a function-evaluation-based technique frequently used in electrical engineering, image and
signal processing, and other fields to address multi-variable minimization problems in which the space
dimensions and the complexity or non-linearity is too high for conventional minimization
algorithms, for instance, those based in derivatives. SA is a generalization of a Monte Carlo
method initially developed for examining the equations of state and frozen states of n-body systems
\citep{metropolis1953}. Later on, \cite{pincus1970, kirkpatrick1983} and \cite{cerny1985}
independently generalized the SA idea to solve discrete optimization problems that involved local
parameter search procedures.

%

The initial idea behind SA consisted on developing an algorithm that would calculate
the (global) internal energy state of a crystalline structure in metallurgic studies
\citep{metropolis1953}.
The goal of the original SA algorithms was to minimize an objective function which resembles
the internal energy of materials. The algorithm took an independent parameter called {\it temperature}
and reduced it following a certain annealing law. For each temperature, a number $N_{gen}$ of
possible {\it test states} were generated from a previous state, and the objective function
was computed for each of those test states; a test state was accepted or rejected based
on the acceptance rules, which depend on the temperature parameter.
If a test state was accepted, it became the current state and, if it was the one with
the minimum absolute energy, it was saved. The temperature was then reduced to begin the
process from the last accepted state. At each temperature, the loop in which the states
are generated and tested for acceptance is commonly called the {\it Metropolis loop}.

The key feature of the SA algorithm is the way in which the moves to a new test state are accepted.
All the test states with a energy (objective function) lower than the current energy are accepted,
but those test states with a higher energy can be accepted or rejected based on a probability
computed from the energy difference between the current and the test state.
This is how SA algorithms avoid stopping at local minima, since a state with a higher
energy configuration, an {\it uphill movement}, can be accepted given a certain probability.

Applied to a general minimization problem, the SA method consists of three functions:
1) a probability density function for a N-parameter space, where N are the parameters to minimize.
This function generates new test states,
2) a probability of acceptance function, which determines whether a given step solution
is accepted or discarded, and
3) the {\it annealing} schedule, which defines how each parameter changes with each iteration step.

Several variations of SA are described in the literature
\citep[see e.~g. ][]{ingber1996, corana1987, dreo2006, otten1989, salamon2002},
but here we focus on the original SA and the so called {\it Fast Simulated Annealing}
\citep[FA, ][]{szu1987} techniques.

The probability distributions of the original SA are generally described as {\it Boltzmann Annealing}
\citep[BA, ][]{metropolis1953,kirkpatrick1983}. Based on the BA notation, the probability of
acceptance function can be expressed as:
\begin{equation}\label{eq:acceptance}
P= exp(-\Delta E / T),
\end{equation}

where $\Delta E$ is the {\it energy} difference between states and T is the schedule of the
annealing, which is denoted by T by analogy with the temperatures of the different energy
states in thermodynamics:

\begin{equation}\label{eq:temperature}
T(k)= \frac{T_0}{\ln(k)},
\end{equation}

In this equation, $T_0$ is the initial temperature, chosen for the algorithm to explore the
full range of the parameters to be searched, and $k=0,1,...,N$ with $N$ large enough to
achieve a convergence. The BA algorithm picks the range of the parameters to be searched and
starts iterations following this temperature law in search for the
'global minimum energy state' of the system. The generating function for the new states is
a gaussian function centered in the current state.

A variation of the SA algorithm is called {\it Fast Simulated Annealing} \citep[FA, ][]{szu1987}
in which the gaussian generating distribution is replaced by a Cauchy distribution. This
lead to a better access to distant states due to the long tail of the Cauchy
distribution, thus improving the exploration of the parameter space in the search for
the global minimum. To ensure the convergence properties of the new algorithm a change in the
schedule of the annealing temperatura is needed, following a faster function $T(k)=T_0/k$, from
which the FA algorithm takes the name. Our code implements a generating distribution
similar to FA, as described in section \ref{subsubsec:statefunction}.


\subsection{Adaptive Simulated Annealing}

The \verb;rvfit; code that we present in this paper to perform fast fitting of
radial velocity curves is based on he Adaptive Simulated Annealing (ASA) algorithm.
ASA \citep{ingber1989,ingber1993,ingber1996,chen1999}, was created with the objective
of speeding up the convergence of standard SA methods. 

ASA has been already applied to the computation of orbits for binary systems with radial
velocities and visual measurements \citep{pourbaix1998}. But the cases where the
only observations avaliable are the radial velocities are many more, e.g. exoplanets and
single/double line eclipsing binaries. In the majority of these cases there are not
relative positional measurements. Also in these cases, even if light curves exist, sometimes
is necessary to fit the radial velocities alone to feed initial sets of parameters
to more elaborated modeling codes. The advantage of this approach is to reach
a full solution for the physical parameters of the binary system faster.
The \verb;rvfit; code that we present in this paper to provide fast fitting of
radial velocity curves is based on the ASA algorithm.

%

The basic structure of the ASA algorithm is the same of the classical SA. There are,
nevertheless, some key differences: new distributions for the acceptance and state
functions and a new annealing schedule; the use of independent temperature
scales for each fitted parameter and for the acceptance function;
and the use of a {\it re-annealing} at specific intervals. We explain each one of those
steps in the following subsections.

\subsubsection{The state function}
\label{subsubsec:statefunction}

Defining U and L as the vectors with the upper and lower bounds of the parameter space, in
which each point is represented by a vector x, for each parameter $x_i$ the
function used to generate new test points is:

\begin{equation}\label{eq:xnew}
  x^{new}_i=x_i+q_i(U_i-L_i),
\end{equation}

where $q_i$ is the following distribution function defined in the interval [-1,1]
and centered around zero

\begin{equation}\label{eq:qi}
  q_i=sgn\left(v_i-\frac{1}{2}\right) T_{i,gen}(k_i)
  \left[ \left( 1+\frac{1}{T_{i,gen}(k_i)} \right) ^{|2v_i-1|}-1 \right]
\end{equation}

In this function, $v_i$ is a random number in the [0,1] interval and $T_{i,gen}(k_i)$ is
the generating temperature for the parameter $x_i$, which depends on $k_i$ called {\it annealing times},
an independent index for each parameter. The function $sgn()$ determines the sign of the
expresion inside the parentheses.

\begin{figure}
\epsscale{.80}
\plotone{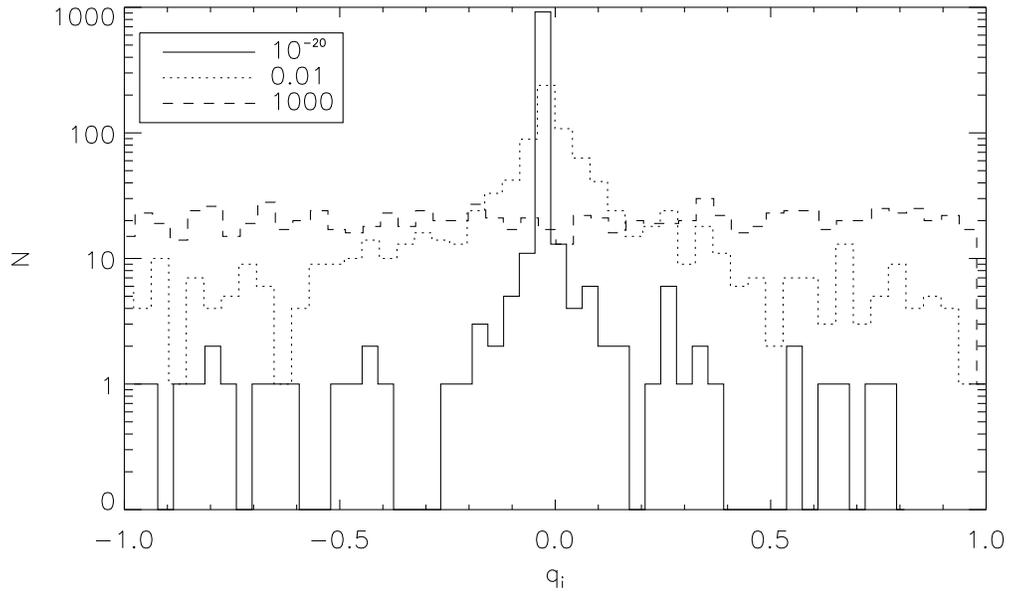}
\caption{Behaviour of the distribution function $q_i$ for three different generating
temperatures $T_{gen}$: dashed for $T_{gen}=1000$, dotted for $T_{gen}=0.01$ and solid
for $T_{gen}=10^{-20}$. The effect of reducing the temperature is to concentrate the
generated points around the central value $q_i = 0$, but still allowing the generation of
some points far away. The parameter $c$ was fixed to 20 as in our code.}
\label{fig:histo_q}
\end{figure}

We show the behaviour of this distribution function in Figure \ref{fig:histo_q} for
three different generating temperatures. The distribution of points given by this function
concentrates around the central value when $T_{i,gen}$ is reduced. But even in the case of very
low temperatures, some points are generated far away from the central point
allowing for a scan of the space parameters and moving away from the central value
whether or not a better solution is found. This avoids local minima.

\subsubsection{The annealing schedules}

For this generating function, the asociated generating temperature follows an exponential function:

\begin{equation}\label{eq:Tigen}
  T_{i,gen}(k_i)=T_{i,gen}(0) e^{-ck_i^{1/n}},
\end{equation}

where $T_{i,gen}(0)$ is the initial value, $k_i$ is a natural number which depends on each parameter
to fit, $n$ is the number of parameters to fit, and $c$ is a constant which depends on each problem and
need to be adjusted. A similar equation is needed for the acceptance temperature:

\begin{equation}\label{eq:Ta}
  T_a=T_a(0) e^{-ck_a^{1/n}}
\end{equation}

Based on the statistical properties of the algorithm \citep{ingber1989}, a global minimum
can be reached with these annealing laws while maintaining a fast convergence rate.
The ASA algorithm becomes {\it adaptive} by the definition of an independent temperature value
for each parameter.

\subsubsection{Re-Annealing}

The last difference between the SA and ASA algorithms is {\it re-annealing}. The idea behind
re-annealing is that the rate of change of the annealing schedule can be changed independently
for each parameter throughout the convergence proccess. In its way towards the global
minimum, the algorithm travels the parameter space through points with very different
local topology, i.e. while some parameters vary rapidly in some regions of the
parameter space, others may vary very slowly in those regions.
By using the same generating temperatures ($T_{i,gen}$)
for all the parameters, it is accepted that the cost function behaves isotropically,
i.e., the topology of the parameters space is nearly the same in all points
and in all directions. This situation results in a waste of computational effort.


Therefore, to optimize computational time, ASA decreases $T_{i,gen}$ along the
directions in which the sensitivity
of the cost function is greater, to allow the algorithm perform small steps, and incresases
$T_{i,gen}$ along the directions with a small sensitivity to allow for large jumps.
ASA adapts its performance by re-scaling the generating temperatures ($T_{i,gen}$) and the acceptance
temperature ($T_a$) every $N_{accept}$ acceptances.

To adapt the algorithm's performance, first the sensitivity of each parameter is updated
based on the local topology of the parameter space. This is done by numerically computing
the derivatives:

\begin{equation}\label{eq:sens}
 s_i=\left|\frac{\partial E(x_{best})}{\partial x_i}\right|=
  \left|\frac{E(x_{best}+\delta_i)-E(x_{best})}{\delta_i}\right|,
\end{equation}

where $\delta_i$ is a small step size in the parameter $x_i$ \citep[see][]{chen1999}.
The new generating temperatures are then computed as:

\begin{equation}\label{eq:Tigen_sens}
  T_{i,gen}(k_i)|_{new}=\frac{s_{max}}{s_i}T_{i,gen}(k_i)
\end{equation}

Likewise, the new acceptance temperature is reset to $T_a(k_a)=E(x_{best})$ and $T_a(0)$
is reset to the value of the last accepted cost function value. With the new $T_{i,gen}$
values the corresponding annealing times $k_i$ and $k_a$ are then re-computed as:

\begin{equation}\label{eq:ki}
  k_i=\Bigg[ \frac{1}{c} \log \Bigg( \frac{T_{i,gen}(0)}{T_{i,gen}(k_i)}\Bigg) \Bigg]^n
\end{equation}

\begin{equation}\label{eq:ka}
  k_a=\Bigg[ \frac{1}{c} \log \Bigg( \frac{T_a(0)}{T_a(k_a)}\Bigg) \Bigg]^n
\end{equation}

and the algorithm resumes with the new computed values.

\section{The {\it rvfit} objective function: Radial Velocity Keplerian Orbits}
\label{sec:objectivefunc}

The model that the \verb;rvfit; code tries to fit is the radial
velocity keplerian orbits equation for the case of a double-lined binary, or a
single-lined binary where the second object can be either another star or a planet.
Assuming gaussian uncertainties in the measurements, the function that our implementation of the
ASA algorithm aims at minimizing is $\chi^2$. In the most general case of a double-lined
binary, the $\chi^2$ function is given by

\begin{equation}\label{eq:chisq2radvel}
 \chi^2 = \sum_{i=1}^{N_i} \left(\frac{v_{calc,i}-v_i}{\sigma_i}\right)^2 + \sum_{j=1}^{N_j} \left(\frac{v_{calc,j}-v_j}{\sigma_j}\right)^2,
\end{equation}

where $N_i$ and $N_j$ are the number of radial velocity measurements of the primary
and secondary components, and $v_i$ and $v_j$, and $\sigma_i$ and $\sigma_j$ are the
measured velocities of each components and their associated uncertainties.
In addition, $v_{calc,i}$ and $v_{calc,j}$ in this equation corresponds to the expected
radial velocities of the primary and secondary components in our physical model
calculated using the keplerian orbit equation for each component, i.e.

\begin{equation}\label{eq:radvel1}
 v_{calc,i} = \gamma + K_1 [\cos{(\theta+\omega)}+ e \cos{\omega}],
\end{equation}
and
\begin{equation}\label{eq:radvel2}
 v_{calc,j} = \gamma + K_2 [\cos{(\theta+\omega')}+ e \cos{\omega'}], 
\end{equation}

where $\omega'=\omega+\pi$, since the argument of the periastron of the secondary component
differs by $\pi$ from the argument of the periastron of the primary, $\gamma$
is the center of mass velocity of the system (systematic velocity) measured from the Sun,
$\theta$ is the true anomaly, $e$ is the eccentricity of the orbit, and $K_1$ and $K_2$
are the radial velocity amplitudes for the two components. 

In equations \ref{eq:radvel1} and \ref{eq:radvel2}, the argument of the periastron of the system, $\omega$,
is a constant, and the true anomaly, $\theta$, is the parameter that varies with time.
$\theta(t)$ is given by the equation:

\begin{equation}\label{eq:trueanomaly}
 \theta(t)= 2 \arctan{\Bigg[ \sqrt{\frac{1+e}{1-e}} \tan{\left(\frac{E(t)}{2}\right)\Bigg]}},
\end{equation}

where E is the eccentric anomaly, which is computed from the mean anomaly $M$ by resolving
the Kepler equation

\begin{equation}\label{eq:keplerec}
 E - e \sin{E} = M
\end{equation}

The mean anomaly $M$ is obtained directly from the orbital period $P$ and the time of periastron
passage $T_p$ of the system:

\begin{equation}\label{eq:meananomaly}
 M = \frac{2 \pi}{P} (t-T_p)
\end{equation}

Thus, for a double-lined binary the set of parameters to minimize for a given dataset is
$[P, T_p, e, \omega, \gamma, K_1, K_2]$. For a single-lined binary, where 
we can only measure the radial velocities of one of the components, the set of
parameters to minimize is $[P, T_p, e, \omega, \gamma, K_1]$.

\subsection{Resolving the Kepler equation}
\label{subsec:keplerec}


The core step in the computation of the objective function $\chi^2$ is the iterative
calculation of equation \ref{eq:keplerec} to obtain the value of the eccentric anomaly $E$.
This is a major step in the evaluation of
the objective function since the Kepler equation must be solved numerically and, if
not properly done, the evaluation of the objective function will be slow thus degrading
the performance of the SA algorithm. Since the SA algorithm explores the full range of
eccentricity and mean anomaly values for a given binary, the calculation of $E$ must be robust
and must converge quickly for all values of $e$ and $M$.

Several routines has been developed in the past to do this in an efficient fashion
\cite[see e.g. the review by][]{meeus1998}.
Well-known classical methods to solve the Kepler equation include the Newton method
\citep{duffett-smith1988,meeus1998}, the Kepler method
\citep{meeus1998}, and the binary
search method \citep{sinnott1985}.


In the first versions of the \verb;rvfit; code we implemented a classical Newton algorithm,
but soon realized that this method can be improved and accelerated. We therefore ran
extensive tests to choose a fast and reliable method to solve the Kepler equation
for any parameter combination. We finally chose the following method, which we call
{\it Newton two-step} method:

\begin{itemize}
 \item We compute an initial value of E using a third order Maclaurin series expansion
  of $E$ and Kepler's equation. This ensures that the initial value of $E$ is close to
  its real value and therefore provides faster convergence.
  The implemented equation is the following:
  \begin{equation}\label{eq:Maclaurin}
    E=M+e\sin M + \frac{e^2}{2} \sin 2M + \frac{e^3}{8} (3\sin 3M-\sin M)
  \end{equation}
  
  \item In a second step, a Newton classical iterative method was used to refine the computed
    initial value to the desired accuracy. The E value previously computed is used to feed this
    iterative method. At each iteration a new value of the eccentric anomaly is computed using
    the equation:

    \begin{equation}\label{eq:Newton_iterac}
      E_{k+1}=E_k+\frac{M+e\sin E_k-E_k}{1-e\cos E_k}
    \end{equation}

    This process is repeated until $|E_{k+1}-E_k|<\epsilon$. In this case the accuracy was set
    to a somewhat high value of $\epsilon = 10^{-10}$ radians to avoid systematic errors.
\end{itemize}

\begin{figure}
\epsscale{1.1}
\plottwo{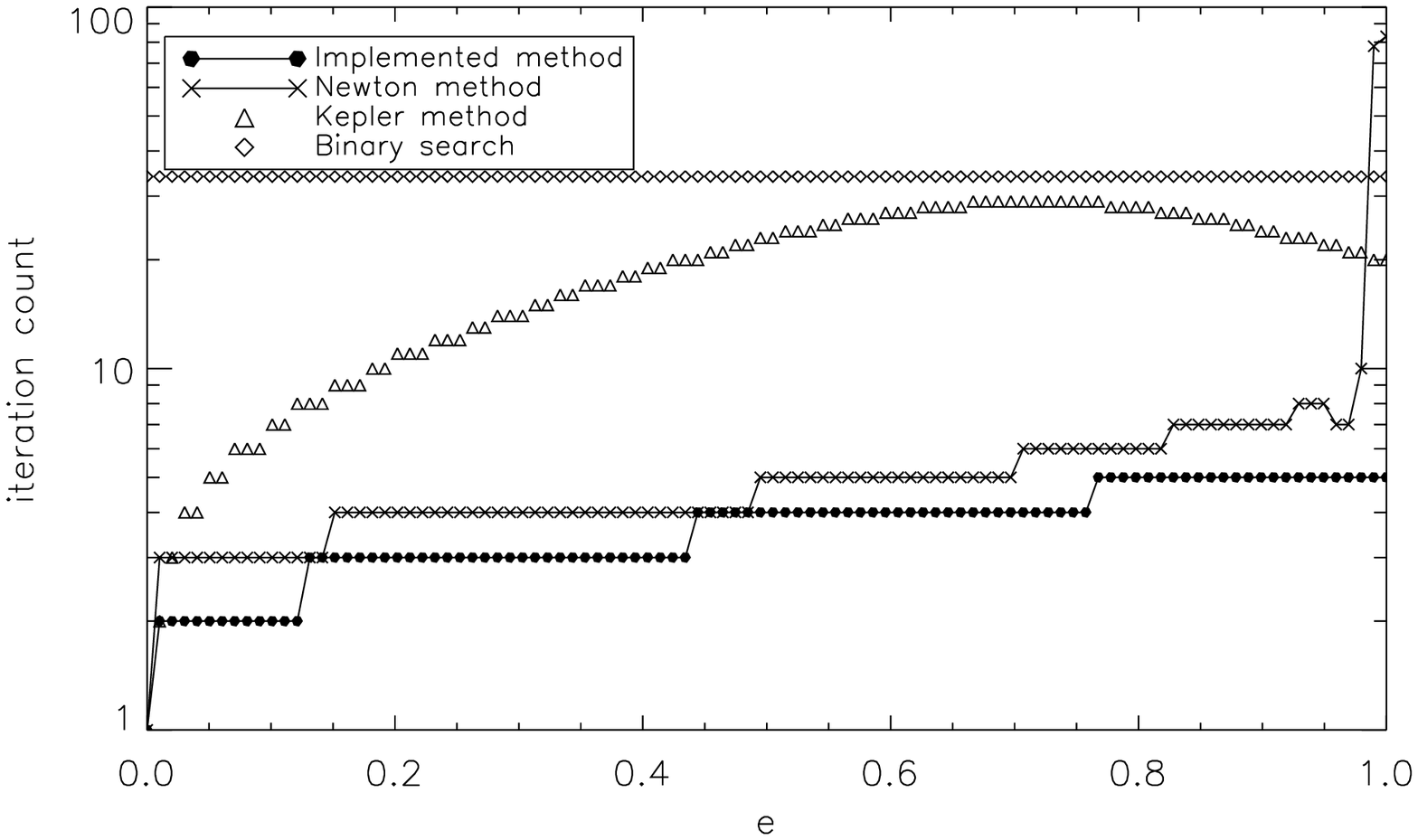}{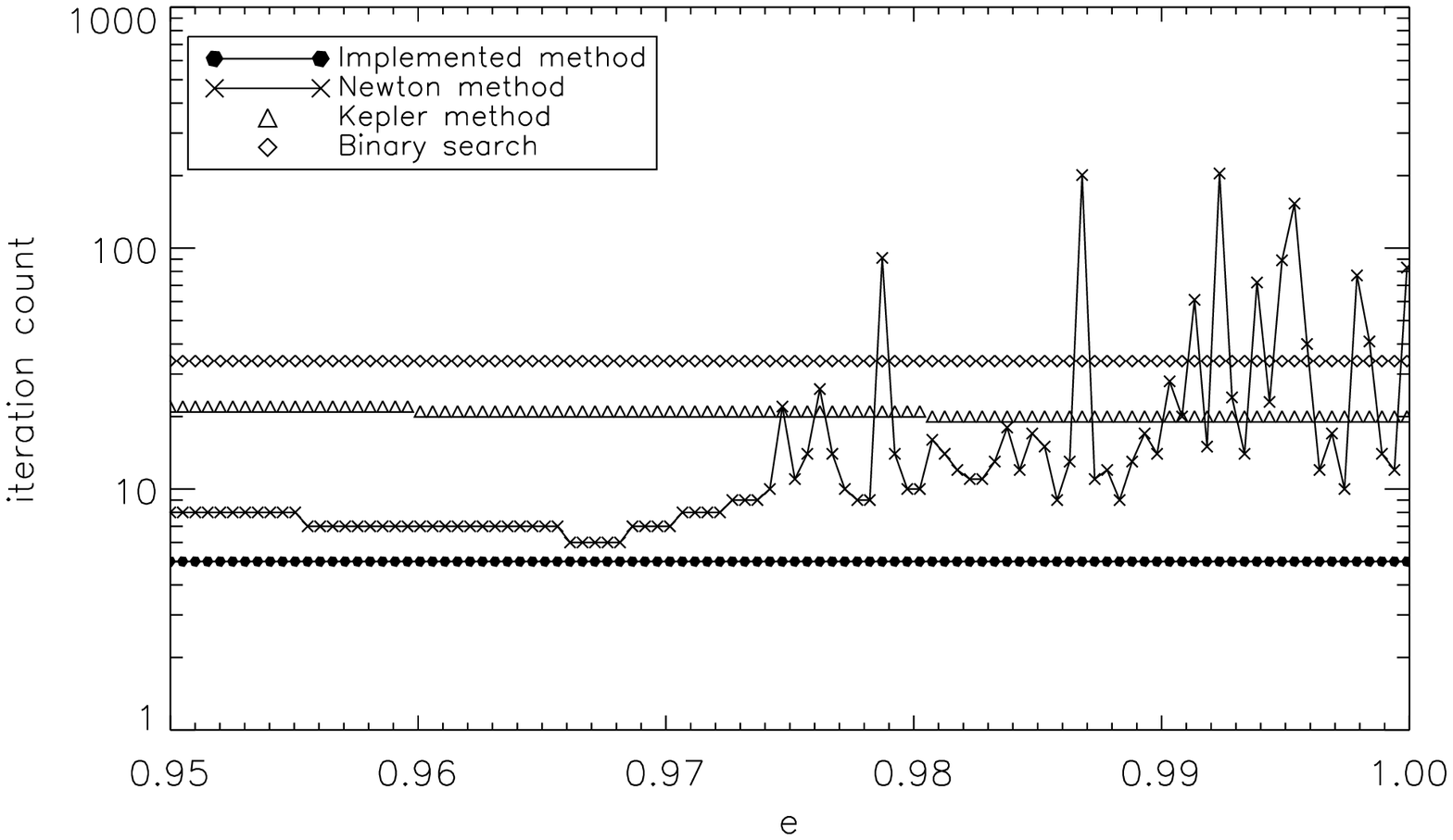}
\caption{Number of iterations versus eccentricity for four different methods to solve
Kepler's equation. The left-side panel shows 100 evenly distributed eccentricity values
between $e=0-0.9999$, for M=0.3 radians. This plot reveals how the classical Newton
method has convergence problems when $e\sim1$.
The right-side panel shows the eccentricity interval $e=0.95-0.9999$, where we compute
100 values for $M=0.3$ radians. Both plots show how our implemented Newton two-step
method is more efficient than all the others.
\label{fig:simulac_keplerec}}
\end{figure}

The number of iterations versus the eccentricity of our Newton two-step routine and other
classical routines is shown in Figure \ref{fig:simulac_keplerec} for an M=0.3 radians
value of the mean anomaly. This value was deliberately selected to show some strange effects
seen in the number of iterations at high eccentricity values for some classical routines.
The discrete jumps in the plots are due to the fact that iteration numbers are integers,
so jumps appears for small values in the logaritmic vertical axis.
The comparison among our two-step method and the other methods are displayed in the
left panel of the Figure \ref{fig:simulac_keplerec} for the full range of eccentricities.

Since we selected 32 steps in the binary search method, the plot of the iteration number
is a constant. In the Kepler method we also implemented the two-step approach mentioned,
computing the initial value of the eccentric anomaly with the eq. \ref{eq:Maclaurin}.
The Kepler method needs from 3 to 30 iterations depending on the eccentricity value
but the behaviour shows a regular trend with a maximun of iterations near e=0.7.
The Newton classical method is very fast, converging to the solution in less than
10 iterations for eccentricities below e=0.95 but shows a strong increment near the
limit of $e<1$. To investigate this behaviour we run aditional tests for the eccentricity
in the range [0.95,0.999] for each method. The results are displayed in
Figure \ref{fig:simulac_keplerec}. A strong random variation is observed in the number
of iterations needed to achieve the desired accuracy, with
peaks above 200 iterations for some values of the eccentricity and less than
20 iterations for the neightbour test. No clear pattern is observed in this variation,
which suggest a weak convergence in this range of eccentricities. The results of the tests
shown here are for M=0.3, but we find a similar behavior for other values of M, affecting
slightly different intervals of eccentricity near the limit of 1.

\cite{meeus1998} has observed this same effect (see his Figures 4 and 5) and
claims that the problem is present for values of M near 0 and eccentricity near 1, i.e. the
parameters that makes the denominator in the equation \ref{eq:Newton_iterac} almost 0.
He quoted some solutions, which are reduced to limiting the value of the
eccentric anomaly correction in the algorithm.

Our implemented Newton two-step algorithm avoids this problem and improves the performance
of the Newton method by starting from a solution very near the correct result,
thus reducing the number of iterations and increasing the stability.

\section{The implemented ASA algorithm}
\label{sec:ASAalgorithm}

In this section we provide a detailed description of our ASA algorithm implementation.
For our implementation we follow the same approach detailed by \cite{chen1999},
except for how we calculate the initial acceptance temperature and other improvements,
as will be described in more detail below. The code was implemented in IDL 7.0.
The ASA code is packed in a main program called \verb;rvfit; but several common routines
are separated in other file for convenience.

Based on the equations described in the previous section, and assuming no previous knowledge
of the orbital parameters of the system, our ASA algorithm minimizes the $\chi^2$ function
in eq. \ref{eq:chisq2radvel} in the multidimensional space given by the following set of
parameters: the orbital period of the system, $P$, the time of periastron passage, $T_p$,
the eccentricity of the orbit, $e$, the argument of the periastron, $\omega$,
the systematic velocity of the system, $\gamma$, and the radial velocity amplitudes, $K_1$ and $K_2$
(only $K_1$ in the case of single-lined binaries and exoplanets).

ASA works by assuming that each measurement is in the form of [$t_i$, $RV_i$, $\sigma(RV_i)$], $i=1,..,N_i$,
where $t_i$ is the time of observation $i$ (in HJD or BJD), $RV_i$ is the measured radial velocity
(m/s or km/s), and $\sigma(RV_i)$ is  the uncertainty of that radial velocity in the same units as RV.
If the binary is double-lined then a set of parameters [$t_j$, $RV_j$, $\sigma(RV_j)$], $j=1,..,N_j$,
for the secondary need to be provided. We assume those uncertainties
follow a gaussian distribution.

%

\subsection{Initialization}
\label{subsec:initialization}

Our algorithm implements some improvements over the one by \cite{chen1999}.
One of those improvements is the initialization of the acceptance temperature ($T_{a0}$).
\cite{chen1999} set $T_{a0}$ from the value of the cost function in an aleatory state.
Instead, we use the \cite{dreo2006} description with the following steps:
1) we compute a set of aleatory perturbations is computed, i.e. 100,
2) we compute the average of the cost function variations due to these perturbations
$<\Delta\chi^2>=<(\chi^2_{k+1}-\chi^2_k)>$,
3) we stablish an initial acceptance probability of $p=0.25$, claimed to be the optimal
value for 5-100 dimensional parameter spaces \citep[see][]{gelman2003,driscoll2006,mackay2006},
and finally 4) we compute the optimum $T_{a0}$ from $p$ and $<\Delta\chi^2>$ as:
\begin{equation}\label{eq:Ta0}
  p=\frac{1}{1+e^{<\Delta\chi^2> / T_{a0}}} \Rightarrow T_{a0} = \frac{<\Delta\chi^2>}{\ln(1/P+1)}
\end{equation}

This change ensures an appropriate initial acceptance temperature $T_{a0}$, optimizing
convergence and therefore reducing computational time.


\subsection{Parameter tunning}

An advantage of the ASA algorithm is that some of the control parameters
are automatically and easily set during the first stages of the execution and are
tuned along the evolution of the temperatures. However, a few of them, namely $N_{acep}$,
$N_{gen}$ and $c$, need to be set by hand and adapted to each particular problem..
Here we explain how we adapt those parameters in \verb;rvfit;.

The $c$ constant (introduced in eqs. \ref{eq:Tigen} and \ref{eq:Ta}) controls the
annealing rate, at which the generating and acceptance temperatures are reduced.
When $c$ is too low the algorithm needs more time to achieve a low temperature stage where
the neightbourhood of the global minimum is explored and refined. So the algorithm slows down.
When $c$ is too high the temperature drops faster and the algorithm can get stuck near
a local minimum depending on the cost function and the parameter space topology.

\cite{chen1999} claim values of $c$ in the range [1, 10] work well. However, our tests
with published observations conclude that $c=20$ provides faster convergence and works well
for the number of parameters we need to fit for. This large value of $c$ makes mandatory
the use of double precision arithmetic in the code variables to avoid computational
overflows or underflows. For some cases, it may be convenient to reduce this value.

We set the other two parameters that need tunning to $N_{gen}=10^4$ and $N_{accept}=10^3$,
an order of magnitude higher than \cite{chen1999}.
These values are justified by the need to have better sampling of the orbital period in cases
where the period of the system is small compared to the range of periods the code explores.
A good example is when short period binaries are found in long observing campaigns.
Of course, these values can be revised for some cases, but we find they work well
for a wide range of synthetic and real radial velocity datasets (see section \ref{sec:test_objects}).

\subsection{Termination}

Another change implemented in our \verb;rvfit; code is the way in which a termination condition
is achieved. A number of stopping rules are described in the literature 
\citep[see ][p. 23]{locatelli2002}, all of them based on the fact that when the algorithm
does not evolve significantly over a number of temperature changes then it must
be stopped. When fitting a model to a set of observations, some aditional
information is provided by the uncertainties, $\sigma$: assuming that the model is
correct and the uncertainties are well computed, then the fitted model must pass through
most of the points at a distance less than $\sigma$. When the algorithm
is stuck near a local minima we can use this criterion wheter it is stopped.

We decided to use a mixed criterion based on the information provided by the uncertainties and 
the progress of the algorithm. Our criterion gathers information about the convergence state
as the algorithm evolves, and with that information and the actual value of the $\chi^2$
makes a decission about stopping. To do this, we make a list with the last $N_\varepsilon$
best values of $\chi^2$ (in our case $N_\varepsilon=5$), and append the best $\chi^2$ to the
list after each temperature step.
When the differences among all the values of $\chi^2$ in the list are less than $\varepsilon$
(with $\varepsilon=10^{-5}$), the annealing loop is stopped if the value of the last $\chi^2$ is:

\begin{equation}\label{eq:chisqtermination}
  \chi^2\leq\sum_{k=1}^{n} \left( \frac{\hat{\sigma}}{\sigma_k} \right) ^2,
\end{equation}

where $\hat{\sigma}=(\sum_{l=1}^{n} \sigma_l)/n$ is the mean of the uncertainties $\sigma_l$
in the data (primary and secondary), with $n=N_i+N_j$. This upper limit is set
assuming that in a good fit some of the measurements are separarated less than $\hat{\sigma}$
from the model.

If this last condition is not met, we reset the acceptance temperature $T_a=T_{a0}$,
and the program resumes the annealing loop.

To prevent the loop from running indefinitely, we impose a limit, $N_{term}$,
the number of reannealings that the annealing loop can do without an improvement in
the $\chi^2$ value. This condition would only happen if the algorithm
got stuck in weird values, when the initial values provided makes the algorithm unstable,
or if the fitted data doesn't correspond to the model.
We impose a high value of $N_{term}=20$ to allow the code to achieve a stable state
before termination.

After terminating the annealing loop, the code computes the uncertainties in the parameters
and uses the parameter values to derive the physical quantities of the system.

\begin{figure}
\epsscale{.90}
\plotone{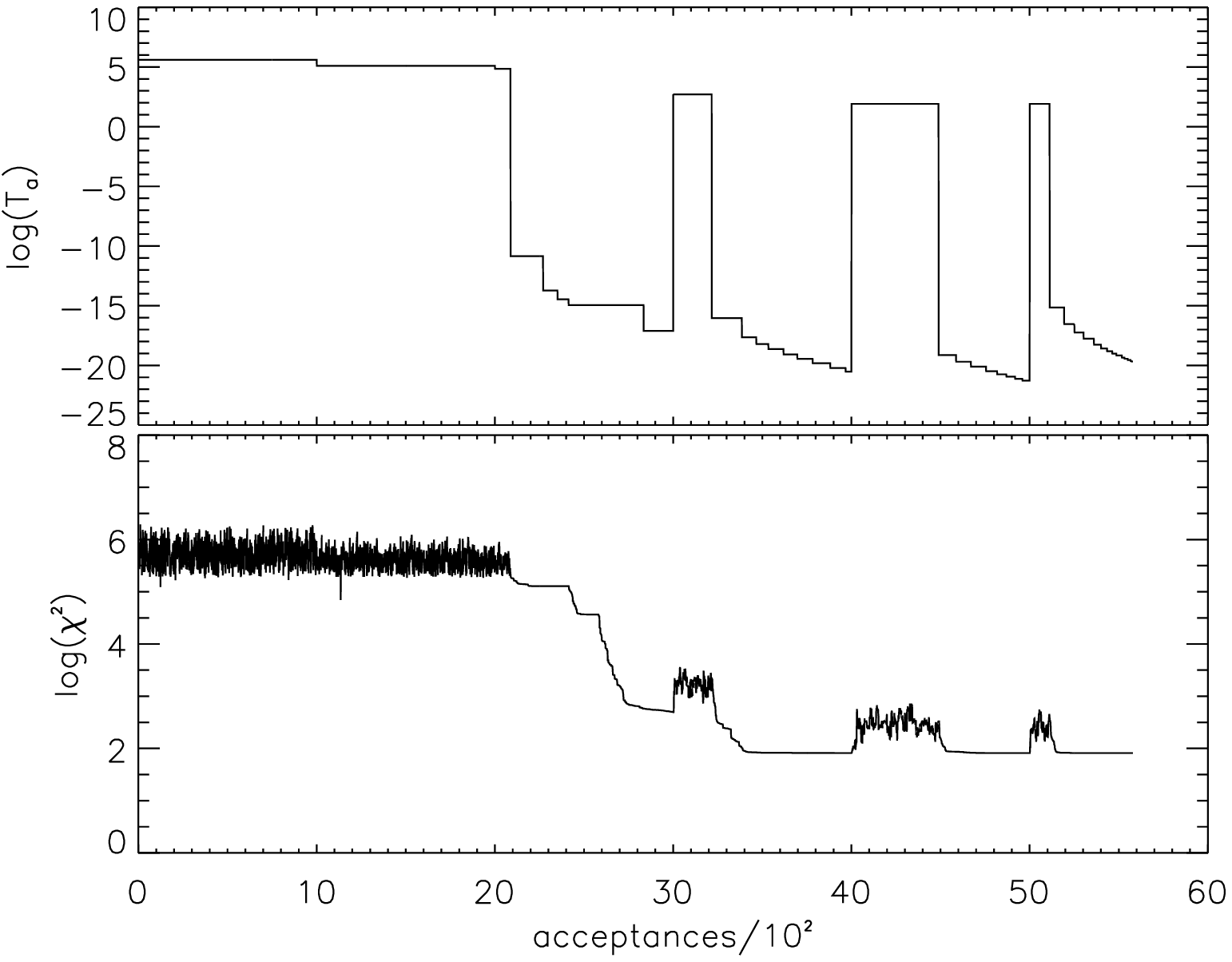}
\caption{The typical behaviour of the acceptance temperature $T_a$ (upper panel)
consists of a sucession of annealings and reannealings. Each upward jump in the temperature
corresponds to a reannealing and it is associated to the exploration of an extended region
in the parameter space, which turns in new acceptances. Those acceptances can be seeing in
the plot of the cost function (lower panel) as upward spikes, since the new acceptances
typically have a higher cost function. This behaviour allows the ASA algorithm to
avoid local minima.}
\label{fig:Ta_F_vs_iterac}
\end{figure}

\begin{figure}
\epsscale{.90}
\plotone{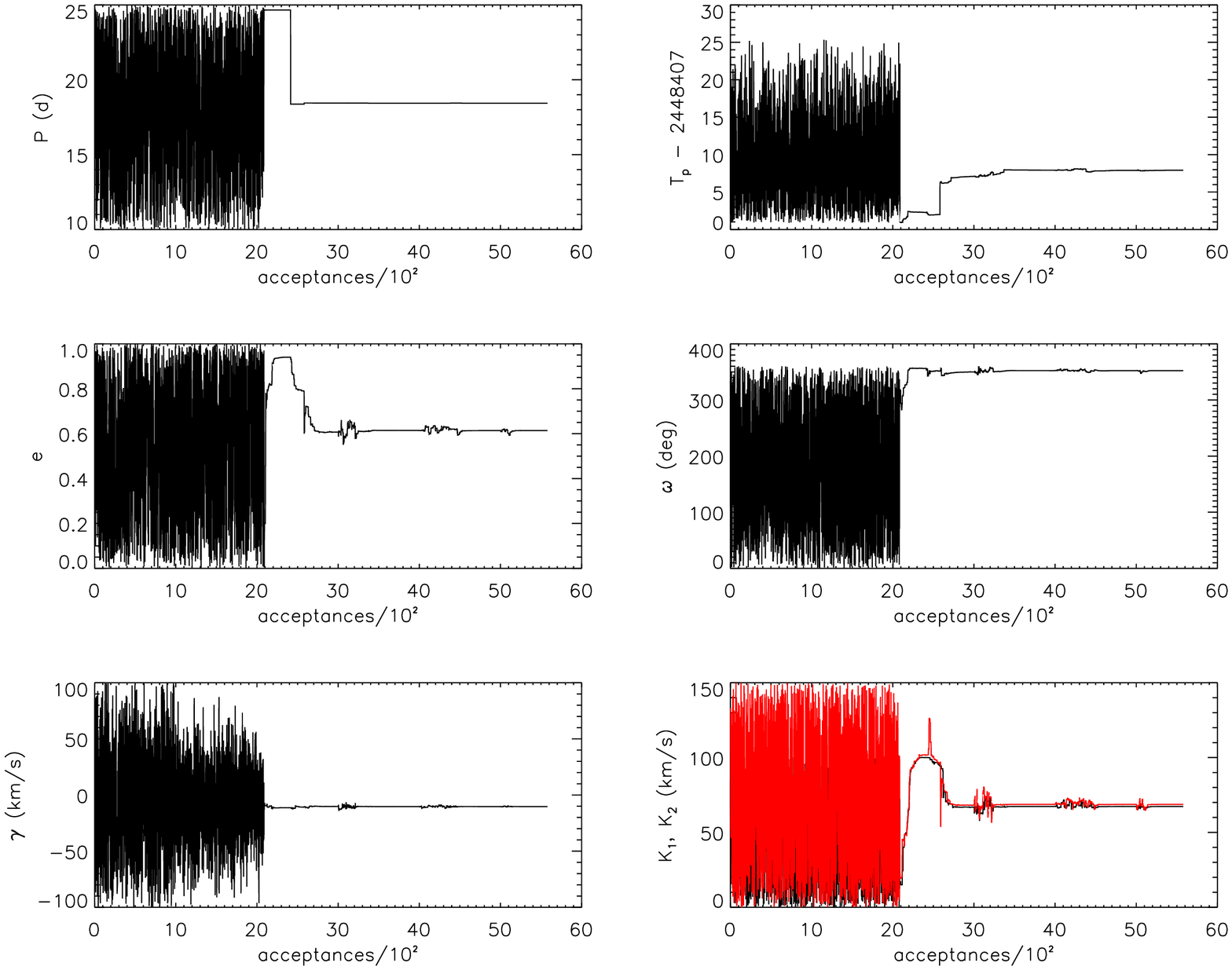}
\caption{Behaviour of the parameter values during a run for the LV Her dataset.
In the amplitudes plot, the black line represents the acceptances for $K_1$ and the red one
for $K_2$.
Note the high density sampling of the parameter domain in the first stages associated
to a high temperatures phase (both $T_a$ and $T_{i,gen}$) in the algorithm.
Once the global minimum region is found, near the acceptance 3400, all the acceptances
are produced in that neightbourhood and the parameters evolution is estabilized.
Also, note that the first parameter to become stable is the period, which show
fluctuations smaller than the thickness of the line in the plot.}
\label{fig:params_vs_iterac}
\end{figure}

The behaviour of the algorithm is shown in Figures \ref{fig:Ta_F_vs_iterac}
and \ref{fig:params_vs_iterac}. Those figures show the acceptance temperature and the
cost function values for all the accepted parameter configurations for a fitting run
to the data of the eccentric double-line eclipsing binary LV Her \citep{torres2009}.
In this fit all the parameters were leave free, including the orbital period.
The figures illustrate the evolution of the acceptance temperature with the annealing
phases of decreasing $T_a$, the subsequent reannealing jumps, and the associated groups
of new acceptances with higher cost function values when a reannealing occurs.
Figure \ref{fig:LVHer_fit} shows the resulting radial velocity curve fit for this binary.

Since ASA is a heuristic method with aleatory generation of test parameter values,
these plots will be somewhat different each time the code is executed. However, they will have
the same overall behaviour of $T_a$ and $\chi^2$ due to the annealing-reannealing cycle.

\section{Computation of uncertainties}
\label{sec:uncertainties}

Once the algorithm has converged to a global minimum solution, \verb;rvfit; computes the
uncertainties of each parameter in one of the two ways described below.

\subsection{Fisher matrix}
\label{subsec:fisher}

The Fisher matrix is a popular procedure to obtain quick uncertainties in the model parameters.
It is based on the central-limit theorem which states that 'a well-behaved
likelihood distribution is asymptotically Gaussian near its maximum'. As a consecuence
this procedure can only describe elliptical uncertainty contours. Their implementation is
straightforward and numerical derivatives can be used if the function doesn't have
analytical form.

Following \cite{andrae2010}, \verb;rvfit; computes the Fischer matrix of all parameters at
the global minimum position as:

\begin{equation}\label{eq:Fishermatrix}
  [F]=\frac{1}{2}
  \begin{pmatrix}
   \frac{\partial^2}{\partial x^2} 		&\frac{\partial^2}{\partial x\partial y}\\
   \frac{\partial^2}{\partial y\partial x} 	&\frac{\partial^2}{\partial y^2}\\
  \end{pmatrix} \chi^2
\end{equation}

The computation of the partial derivatives is done following \cite{coe2009}. Once
the [F] matrix is computed, the covariance matrix is obtained by inversion:

\begin{equation}\label{eq:covFisher}
  [F]^{-1}=[C]=
  \begin{pmatrix}
   \sigma^2_x 	&\sigma_{xy}\\
   \sigma_{yx} 	&\sigma^2_y\\
  \end{pmatrix}
\end{equation}

The main drawback of computing the uncertainties with the Fisher matrix is that it cannot
represent non-linear correlations between parameters, which can be seen in a number
of models.

\subsection{Markov Chain Monte Carlo uncertainties}
\label{subsec:MCMC}

\verb;rvfit; also includes a Markov Chain monte Carlo (MCMC) method for the cases
where more detailed uncertainty distributions $P(x)$ are desired.
\cite{ford2005, ford2006} applied a bayesian analysis to the computation of
uncertainties of exoplanet radial velocity curves. Much of his work can be directly applied
to this problem since the Metropolis-Hastings (MH) algorithm he used to sample
the $P(x)$ distribution uncertainties is the same.

In our case, a separate MCMC code to compute the uncertainties was implemented, using
the simple MH algorithm depicted in \cite{ford2005}. This code works in two basic steps.
First, a new sample $x'$ is generated from a candidate distribution $q(x)$. In our case,
this distribution $q(x)$ is a multidimensional gaussian function of width $\sigma$, which is
a vector composed of the widths for the distributions of each of the fitted parameters.
Second, a decision is made to accept or to reject the new state, using the Metropolis-Hastings
acceptance probability $A$, given by
 
\begin{equation}\label{eq:aceptMH}
  A(x,x')=\min \left( 1, \frac{P(x')}{P(x)} \frac{q(x,x')}{q(x',x)} \right)
\end{equation}

In the cases where the distribution $q(x)$ is symmetric the values $q(x,x')$ and
$q(x',x)$ turn out to be equal and the MH algorithm is the called Metropolis algorithm,
the same one used for the SA algorithm. This is the situation in the typical
case where a gaussian distribution of width $\sigma$ centered in the
actual state $x$ is chosen, as in our MCMC code. Based in the ratio
of likelihoods of the two states, the new acceptance probability is then given by:

\begin{equation}\label{eq:acepM}
  A(x,x')=\min \left( 1, \exp\left[\frac{1}{2}(\chi^2(x)-\chi^2(x'))\right] \right)
\end{equation}

The Markov Chain is built with the accepted values $x'$, if these were accepted,
or with the old $x$, if there is not an acceptance.
The old $x$ {\it must} be appended to the chain whenever there is not
an acceptance. If this is not done, some bias will be introduced in parameters
with a hard bound like $e=0$. See \cite[sec 3.5.3]{eastman2013} for a discussion.

Our MCMC does not make use of a burnin
phase prior to the computation of the Markov Chain because the chain is started
in the parameter values provided by the ASA algorithm which are very near
the maximum of the unknown distribution. But the width $\sigma$ of the proposal
distribution is a key parameter which needs to be tuned up to achieve a good
mixing of the Markov Chain and to obtain a fast sampling of the objective distribution.
This $\sigma$ is usually selected depending of the desired acceptance probability
to achieve a good mixing and we selected again an acceptance probability of $p\sim0.25$,
as in the $T_{a0}$ initialization (see section \ref{subsec:initialization}).
Given that each fitted parameter has its own range and marginalized distribution
the $\sigma$ values must be unique for that parameter.

Our procedure to tune up this vector is as follows. To tune up this vector,
we first set the initial $\sigma$ to 1/3 of each parameter range,
and we measure the acceptance rate for $10^3$ generated states by computing
the individual acceptances using eq. \ref{eq:acepM}. If the acceptance rate is
less than $p\simeq0.25$, $\sigma$ is divided by 1.2 and the process begins
again in the previous point. This process is repeated until the acceptance rate
exceeds the proposed $p\simeq0.25$ value. We are aware that other more elaborated
algorithms exist \citep[see e.g.][sec 3.2]{ford2006}, however this simple
algorithm is fast and does a good job providing a good mixing of the Markov chain.


Once the $\sigma$ for the proposed distribution is computed, the MCMC algorithm
computes the uncertainties for all the parameters. Usually,
a MCMC chain with $10^5$ points is enough to provide a reliable measurement
of the uncertainties, but longer chains might be needed in some cases.

\section{{\it rvfit} perfomance tests}
\label{sec:test_objects}

To test the performance of \verb;rvfit; we compared its results to published solutions for a
set of systems. The systems were selected to cover a range of physical and
observational configurations, i.e. double-lined and single-lined binaries, with
both circular and eccentric orbits and with long and short periods, and exoplanet systems.
We also selected datasets with a variety of observing conditions: with few and many observations,
with high and low signal to noise measurements, and densely and loosely observed.

We do not include datasets aimed at observing specific effects, such as the Rossiter-McLaughlin effect
\citep[see ][]{ohta2005,gimenez2006}, or relativistic or tidal effects \citep[see ][]{sybilski2013}.

We selected two double-line spectroscopic
and eclipsing binaries, two single-line spectroscopic binary stars, and two exoplanet systems.
The published solutions for those systems obtained from the literature and the results of our
\verb;rvfit; code are shown in Table \ref{tab:fittedobjects}. 
All uncertainties were computed using a MCMC chain with $n=10^5$ samples, except for GU Boo
with all parameters free ($n=10^6$ samples) and HD 37605 ($n=5\times10^5$ samples).

\subsection{LV Her}

{\it LV Her} is a double-line eccentric eclipsing binary star \citep{torres2009}. We selected this
system because its high eccentricity, the high quality of its radial velocity curve, the
long orbital period, and the long observing time interval relative to the period.
For this system two fits were done: the first one leaving all the parameters free and the second one
fixing $P$ to the value from the light curve to mimic the fit by \cite{torres2009}.

\begin{figure}
\epsscale{.70}
\plotone{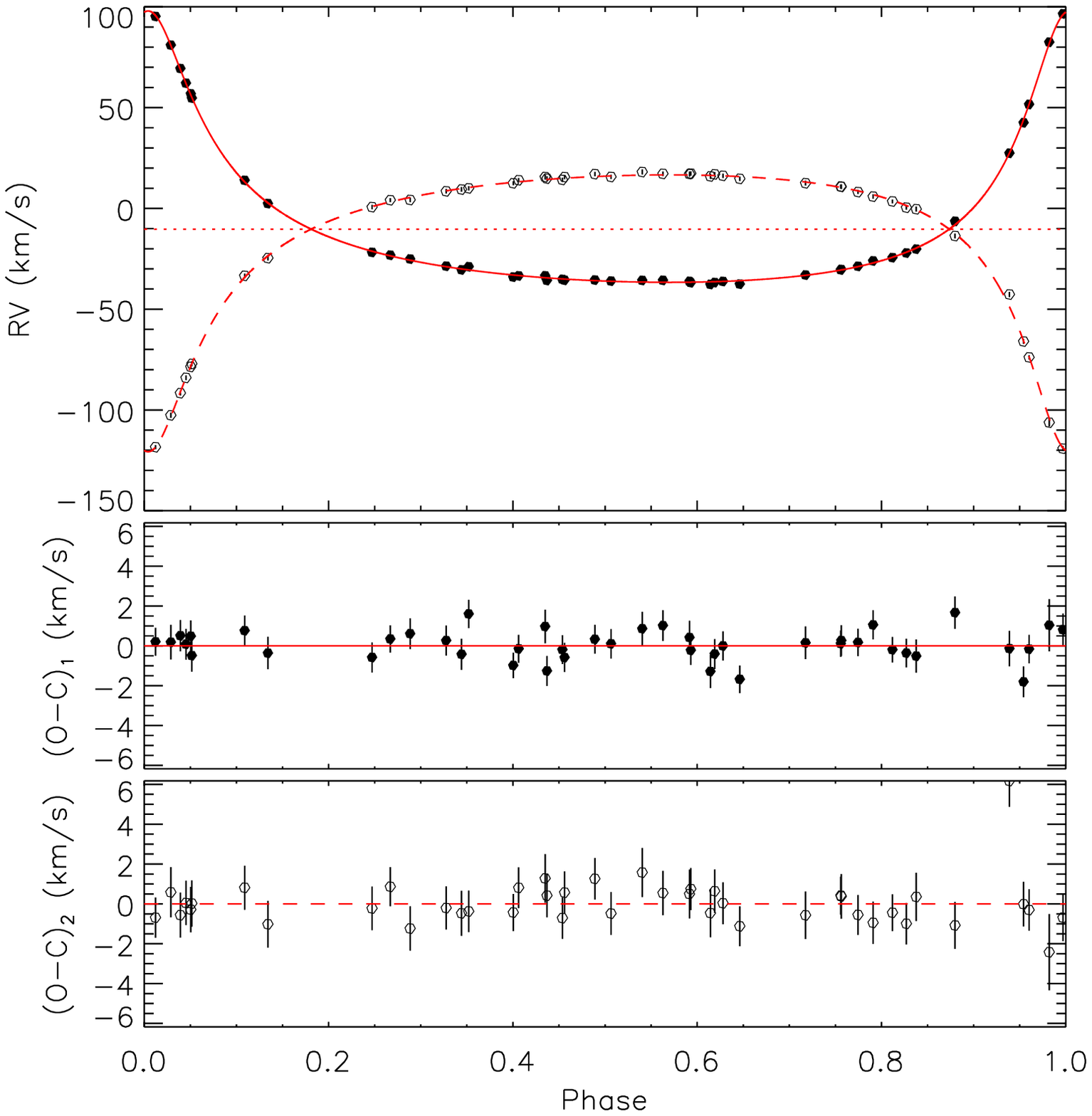}
\caption{Radial velocity curve for LV Her for the fit with all the parameters free,
including the period. The error bars are smaller than the point size in the RV plot.
This solution is visually indistinguishable from the one with $P$ fixed.}
\label{fig:LVHer_fit}
\end{figure}

Our first fit arrives to a solution remarkably similar to the published one
(see Figure \ref{fig:LVHer_fit}).
Our fitted period has a greater uncertainty since the one published was derived from the
photometric eclipses of the system, which are high quality time marks.
Even so, our fitted parameter set, based only in the radial
velocity data, recovers perfectly the overall solution.

Fixing the orbital period to the value from the light curves \citep{torres2009} yields,
again, similar parameters to the ones published.

%

\subsection{GU Boo}

{\it GU Boo} is a double-line and eclipsing binary star \citep{lopez-morales2005},
composed of two nearly equal low-mass main sequence stars ($M_{1,2} \simeq 0.6 M_\odot$)
in an circular orbit with a short orbital period of 11.7 hours.

\begin{figure}
\epsscale{.70}
\plotone{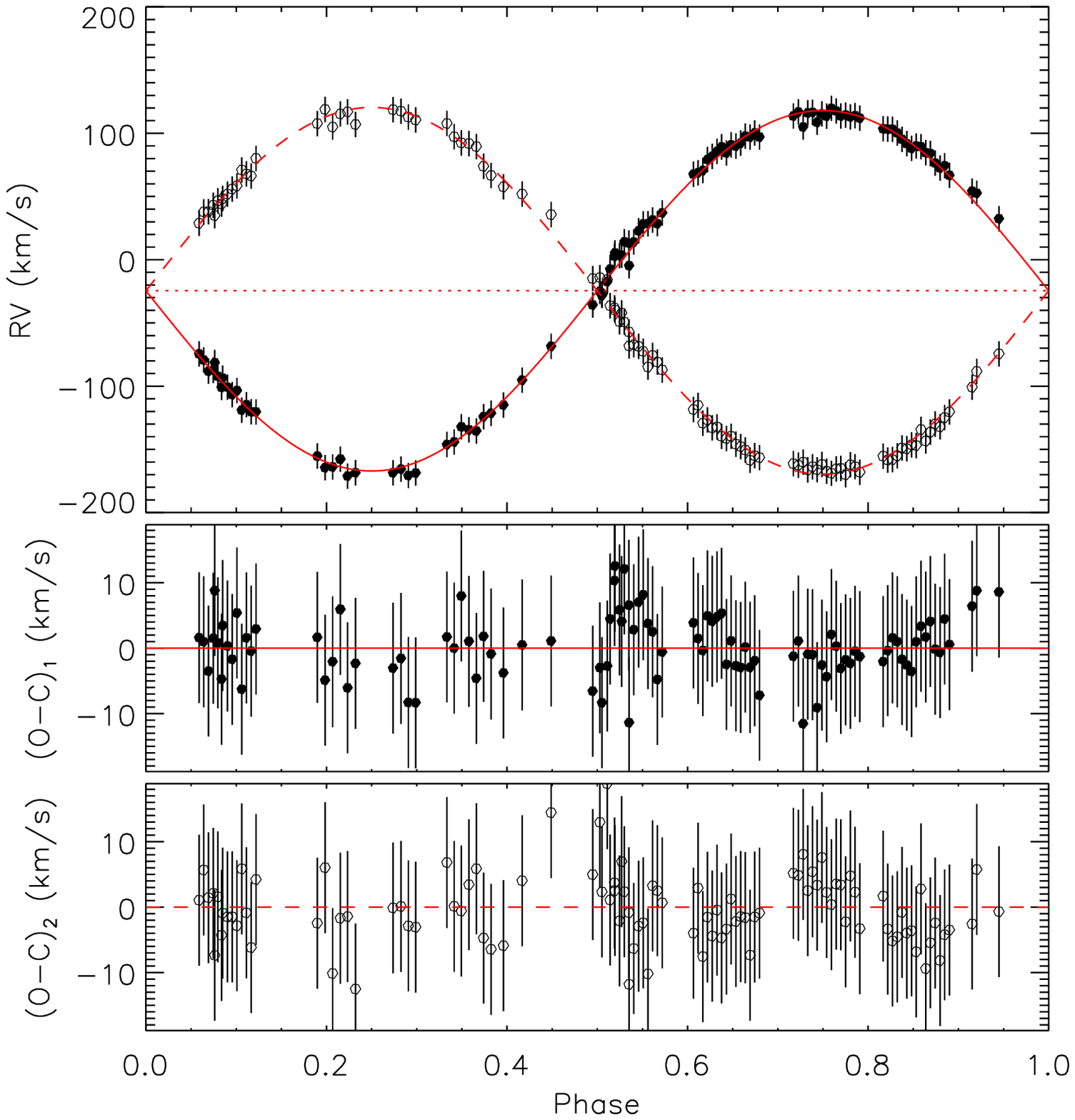}
\caption{Radial velocity curve for the low-mass eclipsing binary GU Boo.
This plot correspond to the second fit, with a circular orbit and in which
we fixed $P$ and $T_P$ to the values obtained from the photometry.
The fit with all the parameters free is indistinguishable from this plot.}
\label{fig:GUBoo_fit}
\end{figure}

Leaving all the parameters free the eccentricity obtained with \verb;rvfit; is compatible with
zero and the uncertainty in $\omega$ is fairly large. This suggest a circular orbit.
The MCMC run shows fairly large uncertainties in those parameters related with
the periastron: the marginalized histograms for $T_p$, $e$, and $\omega$ are severely spread
over a wide range and those of $T_p$ and $\omega$ are multimodal and runing a longer
MCMC chain doesn't fix the situation. This behaviour is a consequence of
the degeneracy in $T_p$ and $\omega$ due to the null eccentricity and is a excellent indicator
of this situation. In Table \ref{tab:fittedobjects} we quote for $P$ the uncertainty
derived from a gausian fit to the histogram and for the other parameters the uncertainties
and values derived from the maximum of the histogram and the
68.3\% shortest confidence interval. We show the fitted RV curve and the residuals
in Figure \ref{fig:GUBoo_fit}

In the second fit we fixed $P$, $T_p$, $e$ and $\omega$ to the values derived by
\cite{lopez-morales2005} from the system's light curves. $\omega$ was fixed to 90 degrees
to match the time of periastron with the instant of the primary eclipse.
This leaves only three parameters free which were easily recovered in our fit.
The uncertainties are gaussian-like and larger than the published values since in the
original paper they were computed by simple error progagation.

\subsection{GJ 1046}

\begin{figure}
\epsscale{.70}
\plotone{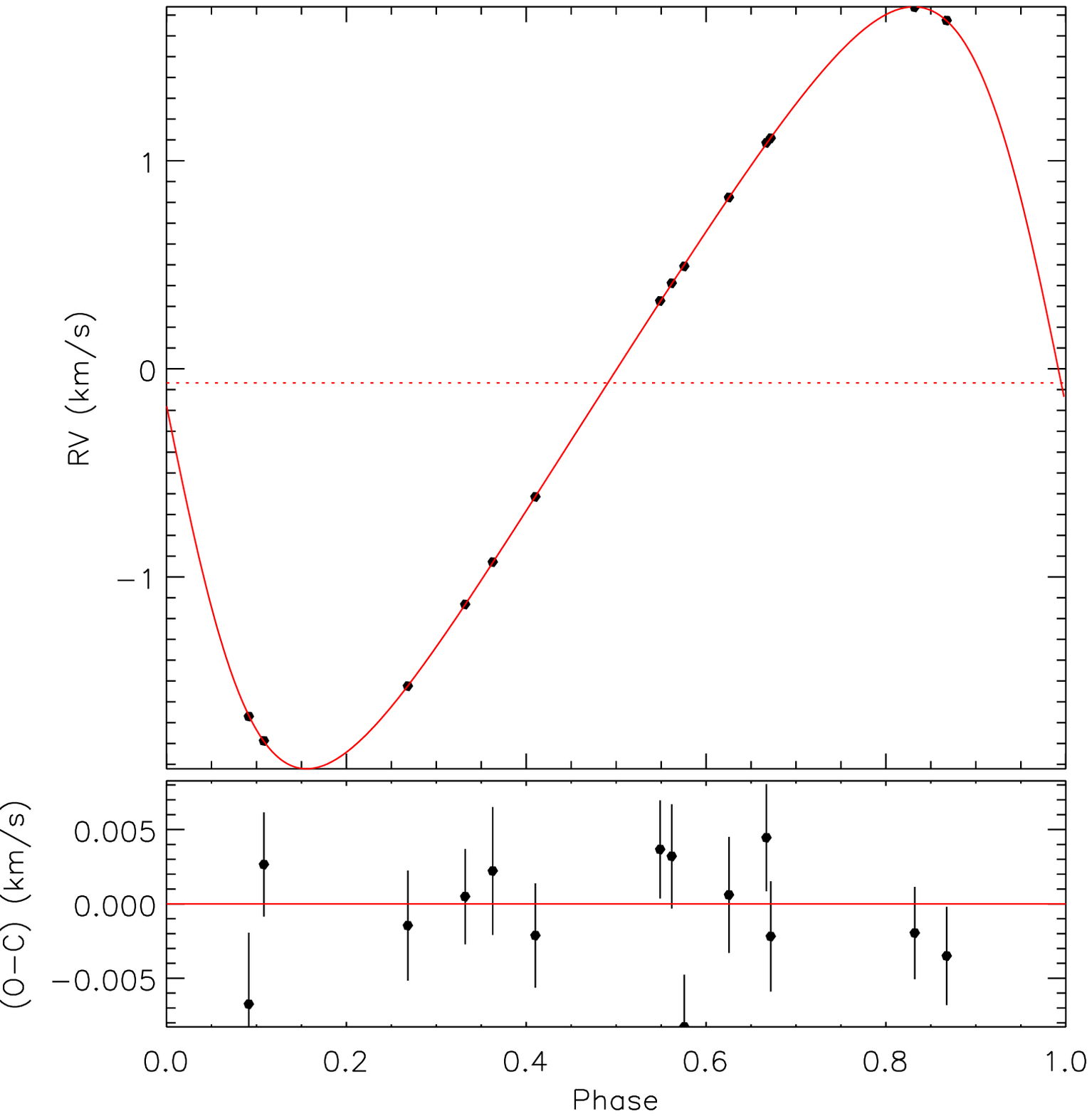}
\caption{Fitted radial velocity curve for the single-lined binary GJ 1046. For this
plot all the six parameters of the Keplerian model were leave free, including
the period. The error bars in the RV curve are smaller than the symbols of the RV values.}
\label{fig:GJ1046_fit}
\end{figure}

{\it GJ 1046} is a system containing a brown dwarf in an eccentric orbit around an
M dwarf \citep{kurster2008}. The system is single-lined because of the low luminosity
of the brown dwarf secondary. We fitted all the parameters simultaneously and obtained
remarkable agreement with the published values (see Figure \ref{fig:GJ1046_fit}),
obtaining a smaller value for the $\chi^2$ than the published value.
The MCMC uncertainties are very well modeled by gaussians and in Table \ref{tab:fittedobjects}
we quoted as uncertainties the $\sigma$ of those gaussians.

\subsection{HR 3725}

\begin{figure}
\epsscale{.70}
\plotone{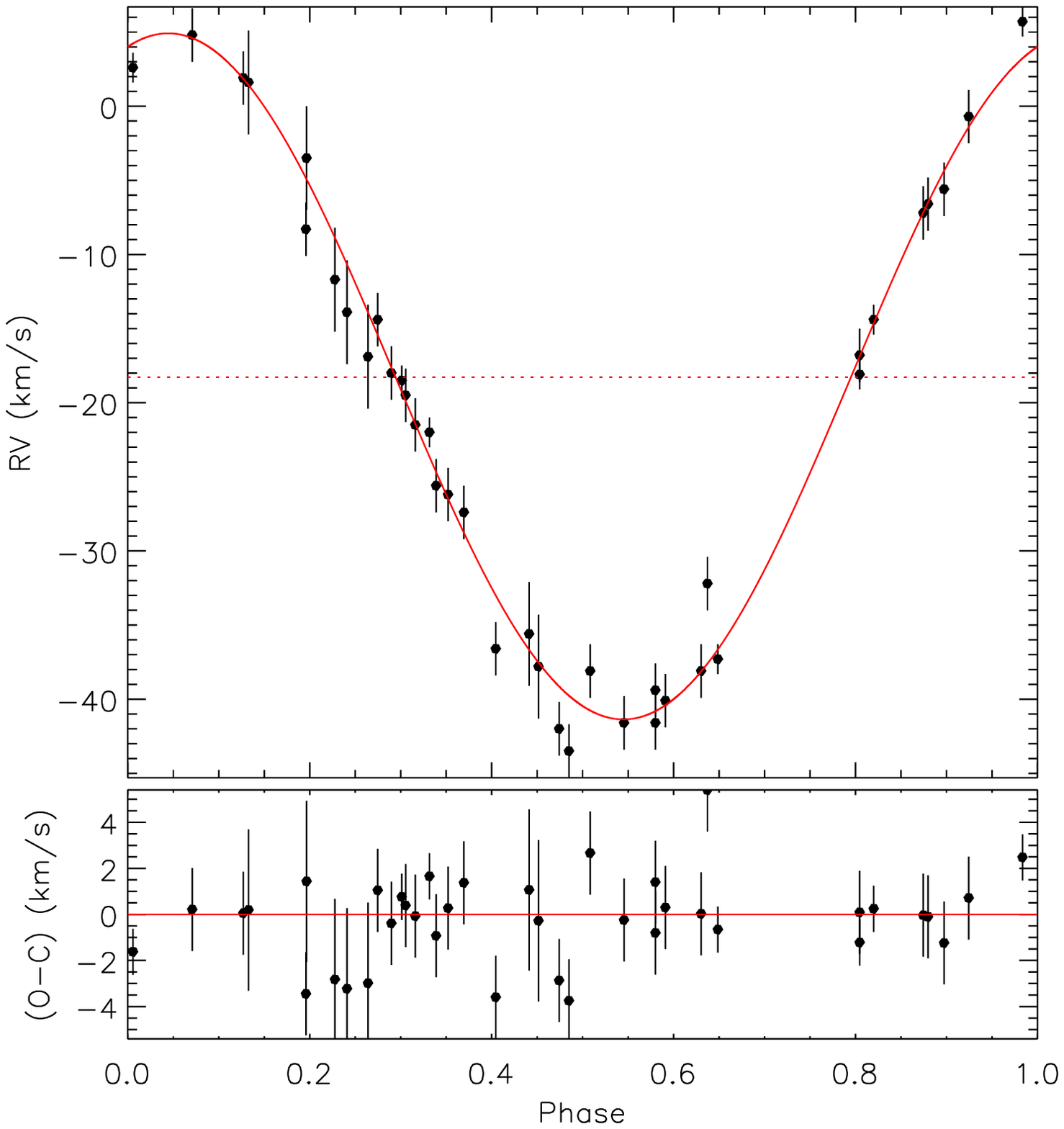}
\caption{Radial velocity curve for HR 3725, a single-line spectroscopic binary with
a circular orbit. This plot correspond to the fit with all the parameters
free, including the period.}
\label{fig:HR3725_fit}
\end{figure}

{\it HR 3725} is a single-lined spectroscopic binary containing a G2 III primary
\citep{beavers1985}. Presumably, the secondary is a dwarf star with a late spectral
type in a circular orbit. Using the published dataset we fitted two models:
the first one with all the parameters free, which confirms the hipothesis of a circular
orbit. In the second model we fixed $e=0$ and $\omega=90$, as the first model suggest.
In both cases, our period is smaller than the published value, which was obtained
by an independent algorithm not stated in \cite{beavers1985}.
The $K_1$ value agrees within uncertainties. Our fitted RV curve is shown in
Figure \ref{fig:HR3725_fit}.

\subsection{HD 37605}

\begin{figure}
\epsscale{.70}
\plotone{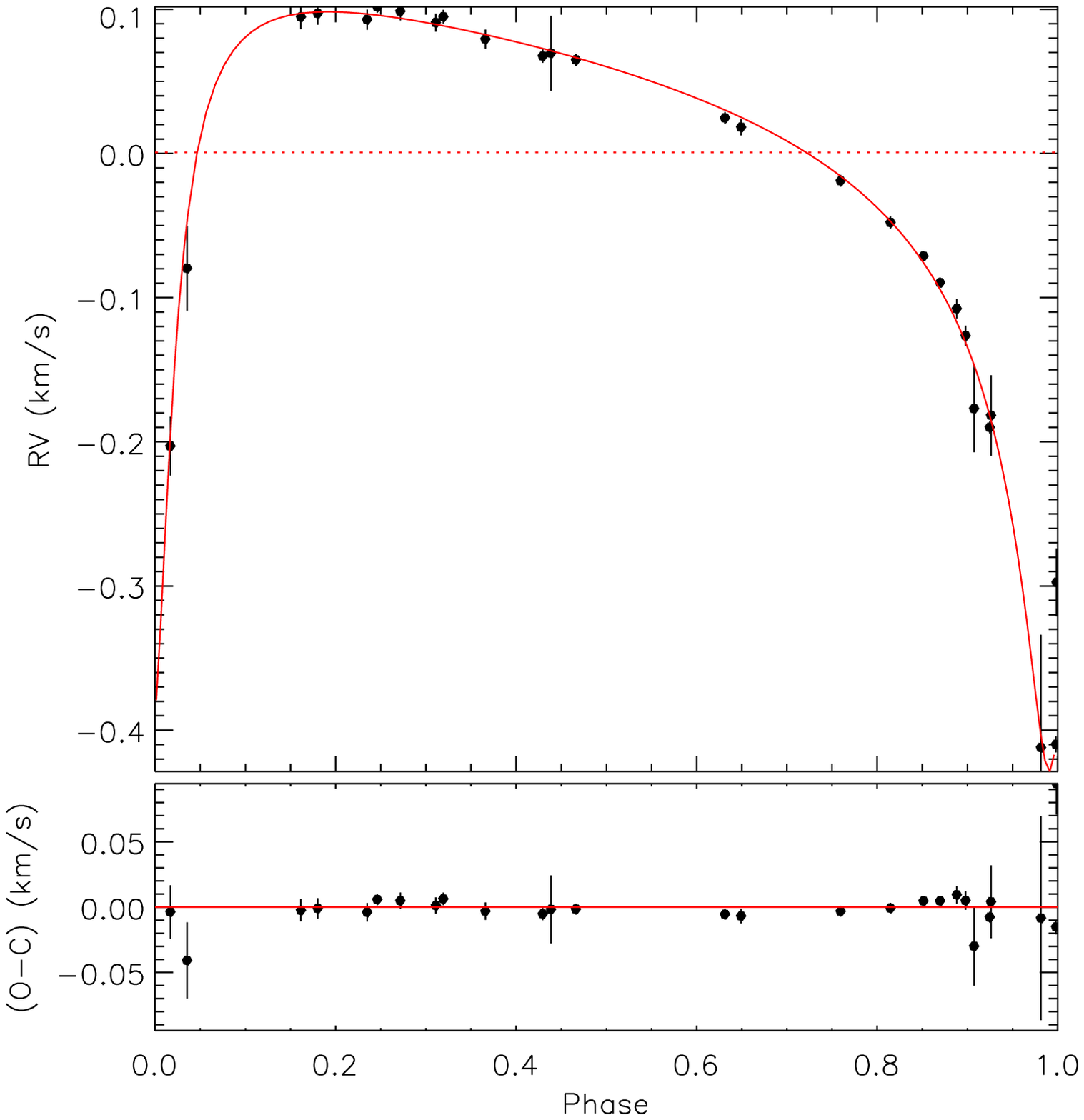}
\caption{Radial velocity curve for the exoplanet system HD 37605 in which we fitted all
the parameters including the orbital period. The plot of the fit with $\gamma=0$
is identical to the eye.}
\label{fig:HD37605_fit}
\end{figure}

{\it HD 37605} was the first exoplanet discovered by the Hobby-Eberly
Telescope \citep{cochran2004}. The planet has a very eccentric orbit ($e=0.737$)
with a period of 54 days and a mass of $m \sin i=2.84$ $M_J$.
Again, we did two fits, one leaving all the parameters free, and another fixing
$\gamma=0$ to mimic the published fit. For both models, the resulting
parameters are in good agreement with the published orbital solution.
The fitted RV solution is shown in Figure \ref{fig:HD37605_fit}.

For this system, the MCMC was run over $5\times10^5$ times,
due to the asymmetry in the marginalized histograms for the
model with all the parameters free. 

%
%

\subsection{Kepler 78b}

\begin{figure}
\plotone{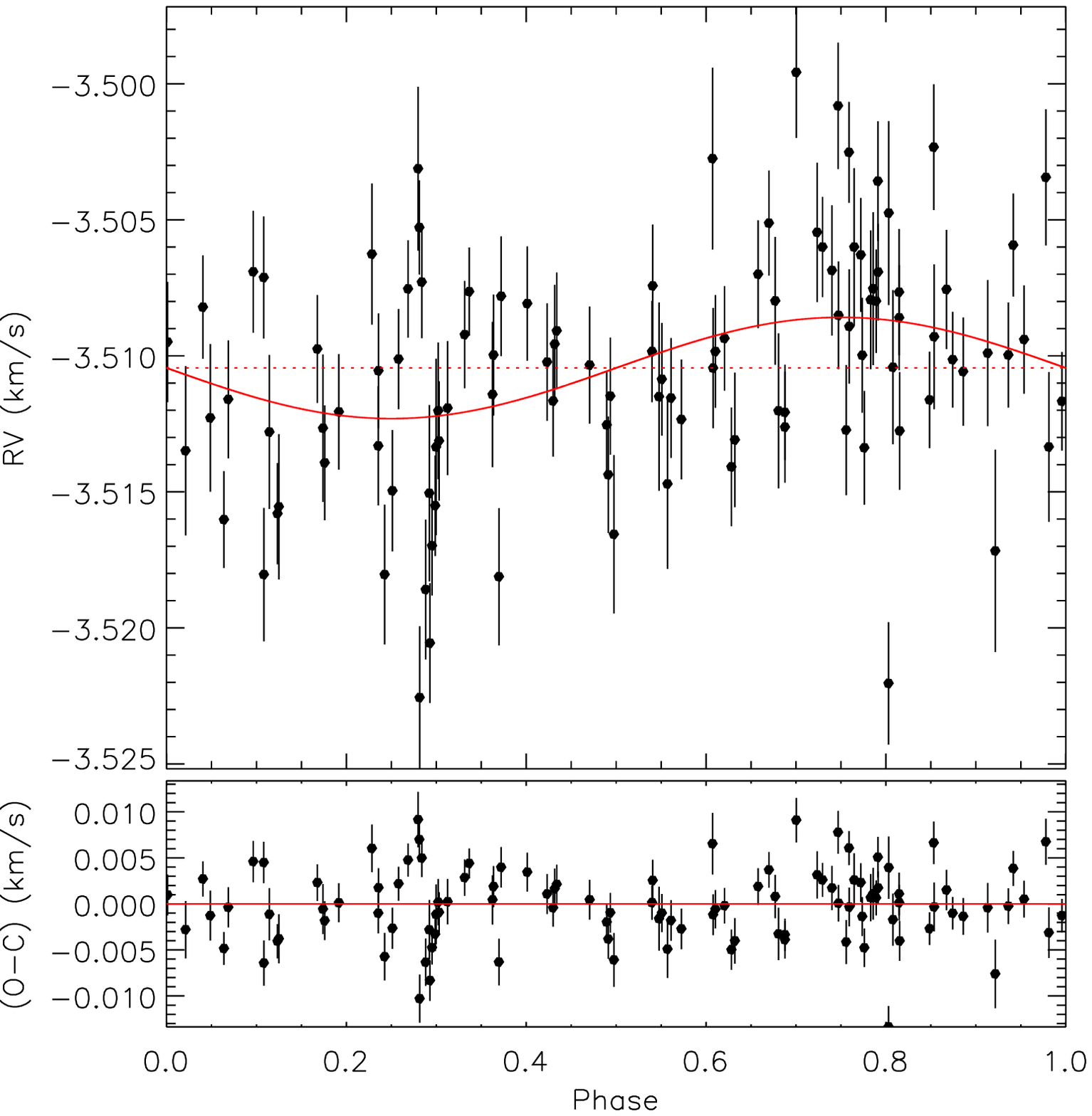}
\caption{Radial velocity curve for the exoplanet system Kepler 78b.
This plot corresponds to the solution with circular orbit and with
$P$ and $T_P$ fixed from the Kepler photometry.}
\label{fig:K78b_fit}
\end{figure}

{\it Kepler 78b} is the first Earth-like transiting exoplanet discovered in the Kepler mission data
\citep{pepe2013, sanchis-ojeda2013, howard2013}. It orbits its parent solar-like star in a
nearly circular orbit and, due to their small mass-ratio, the HARPS-N dataset \citep{pepe2013}
has a very low signal to noise ratio (see Figure \ref{fig:K78b_fit}).
So, this system poses a very stringent test to our code since it must recover the very
low-amplitude RV signal caused by this exoplanet.

In a first fit, we kept $T_p$, e, $\omega$ fixed following a similar approach to \cite{pepe2013}.
$T_p$ was set to the reference epoch in their Table 1, the eccentricity was set to 0
and $\omega$ was set to 90 degrees to match the instant of the transit with $T_p$.
Note that they fitted the mean longitude $\lambda_0$ at epoch $T_0$.
Our fit results in slightly lower values of $P$ and $K_1$, near the limit of uncertainties.
In a first test we set a period range between 0.1 and 1.0 days, but the algorithm
found solutions with
$P=0.265024$ ($\chi^2$=327.73) and
$P=0.748368$ ($\chi^2$=330.62),
both with lower $\chi^2$ than the Kepler period ($\chi^2$=339.98).
The presence of those solutions led us to constrain the period search to the
interval [0.3, 0.5] days to match the period of the planet from Kepler data,
since this is a local minimum in a wider period domain.

Since there is a precise period computed by \cite{sanchis-ojeda2013} from the Kepler
mission photometry we ran another fit but now fixing P, e, and $\omega$.
The value for $K_1$ has a better agreement with the value
published in \cite{pepe2013}. This is due to the small change in the
period which varies the overall solution, as shown in the figure.

%

\begin{sidewaystable}
\begin{center}
{\tiny
\begin{tabular}{ccccccccccc}
\hline
\hline
Object Name	&$P$		&$T_p$		&$e$		&$\omega$	&$\gamma$	&$K_1$		&$K_2$		&$\chi^2$	&Remarks	&Reference\\
		&(d)		&(HJD or BJD)	&-		&(deg)		&(km/s)		&(km/s)		&(km/s)		&		&		&\\
		
\hline
\hline
\multicolumn{11}{c}{Double-line binary systems}\\
\hline
LV Her		&18.4359535$^f$	&2453652.19147$^f$&0.61273(73)	&352.20(.24)	&-10.278(94)	&67.24(.19)	&68.59(.27)	&-		&-		&[1]\\
		&18.43600(14)	&2448414.904(29) &0.6137(19)	&352.22(.27)	&-10.291(88)	&67.31(.24)	&68.67(.33)	&81.416230	&all free	&[this work]\\
		&18.4359535$^f$	&2448414.9117(61)&0.6138(20)	&352.23(.28)	&-10.288(94)	&67.32(.26)	&68.68(.35)	&81.498606	&$P$ fixed	&[this work]\\
\hline
GU Boo		&0.4887280$^f$	&2452723.9811$^f$ &0.0$^f$	&-		&-24.57(.36)	&142.65(.66)	&145.08(.73)	&-		&-		&[2]\\
		&0.488877(73)	&2452724.033($_{-12}^{+16}$)
						  &0.011($_{-11}^{+05}$)
								&137($_{-09}^{+14}$)
										&-24.81($_{-.61}^{+.78}$)	
												&142.5($_{-1.6}^{+1.0}$)	
														&144.8($_{-1.7}^{+1.2}$)	
																&46.712033	&all free	&[this work]\\
		&0.4887280$^f$	&2452723.9811$^f$ &0.0$^f$	&90$^f$		&-24.64(.69)	&142.4(1.3)	&144.9(1.5)	&51.254343	&$P$, $T_p$, $e$, $\omega$ fixed
																				&[this work]\\

\hline
\multicolumn{11}{c}{Single-line binary systems}\\
\hline
GJ 1046		&168.848(30)	&2453225.78(.32)&0.2792(15)	&92.70(.50)	&-		&1.8307(22)	&not aplicable	&12.7		&-		&[3]\\
		&168.845(21)	&2453225.78(.30)&0.2792(12)	&92.69(.49)	&-0.0672(21)	&1.8307(20)	&not aplicable	&12.466580	&all free	&[this work]\\
\hline
HR 3725		&66.717(4)$^f$	&2444667.86(.25)&0.0$^f$	&-		&-18.93(.38)	&23.30(.52)	&not aplicable	&-		&-		&[4]\\
		&66.6683($_{-62}^{+43}$)
				&2444665($_{-04}^{+26}$)
						&0.002($_{-02}^{+18}$)
								&344($_{-24}^{+143}$)
										&-18.27($_{-.30}^{+.27}$)
												&23.14($_{-.44}^{+.46}$)
														&not aplicable	&45.036888	&all free	&[this work]\\
		&66.6673(47)	&2444617.93(.16) &0.0$^f$	&90$^f$		&-18.28(.26)	&23.08(.47)	&not aplicable	&45.056023	&$e$, $\omega$ fixed	&[this work]\\
\hline
\multicolumn{11}{c}{Exoplanet systems}\\
\hline
HD 37605	&54.23(.23)	&2452994.27(.45)&0.737(10)	&211.6(1.7)	&-		&0.2629(55)	&not aplicable	&-		&-		&[5]\\
		&54.30($_{-.32}^{+.14}$)
				&2452939.82($_{-.65}^{+.72}$)
						&0.7351(87)	&211.1(1.4)	&0.0002(16)	&0.2618($_{-56}^{+54}$)
														&not aplicable	&42.753639	&all free	&[this work]\\
		&54.21(.17)	&2452939.96(.51) &0.7342(78)	&211.0(1.1)	&0.0$^f$	&0.2623(49)	&not aplicable	&42.326212	&$\gamma$ fixed	&[this work]\\
\hline
Kepler 78b	&0.3550(4)	&2456465.076392$^f$&0.0$^f$	&-		&-3.5084(8)	&0.00196(32)	&not aplicable	&-		&-		&[6],[7],[8]\\
		&0.34951(15)
				&2456465.076392$^f$&0.0$^f$	&90$^f$		&-3.51035(21)	&0.00173($_{-38}^{+24}$)
														&not aplicable	&339.98215	&$T_p$, $e$, $\omega$ fixed&[this work]\\
		&0.35500744(6)$^f$ &2456464.1575(98) &0.0$^f$	&90$^f$		&-3.51044(21)	&0.00186(28)	&not aplicable	&326.59471	&$P$, $e$, $\omega$ fixed&[this work]\\
\hline
\end{tabular}
\caption{\footnotesize Results for the fitting of radial velocities data for the selected systems.
All the uncertainties were computed using a MCMC chain (see text). For each parameter, the symmetric
uncertainties were computed by fitting a gaussian function to the marginalized histogram.
In this case the quoted uncertainty is the $\sigma$ of the fitted gaussian function.
The asymmetric uncertainties were computed using the 68.3\% shortest confidence interval over the
histogram. In all the cases, the uncertainties are expressed as the two last digits of the parameter,
with the decimal point present where necessary for guidance.
$^f$ means fixed and adopted from other section in the original paper.
The references are:
[1]=\protect\citet{torres2009} (there is a typo in the published value of $\gamma$ [Torres, priv. comm.]),
[2]=\protect\citet{lopez-morales2005},
[3]=\protect\citet{kurster2008},
[4]=\protect\citet{beavers1985},
[5]=\protect\citet{cochran2004},
[6]=\protect\citet{sanchis-ojeda2013},
[7]=\protect\citet{pepe2013},
[8]=\protect\citet{howard2013}}
\label{tab:fittedobjects}
}
\end{center}
\end{sidewaystable}


\section{Discussion}
\label{sec:discussion}

\subsection{Treatment of circular orbits}
\label{subsec:circular}

As shown in section \ref{sec:test_objects} when the eccentricity is null the periastron remains
undefined, so the parameters related can be set to a convenient value. In eclipsing binary stars or
transiting exoplanets with circular orbits it is common to define the periastron as the point
of the inferior conjunction since it is well constrained by the center of the main eclipse
($T_0$) and can be used as the phase reference. So it is common to fix $T_p=T_0$ and
$\omega=\pi/2$ to match the defined line of nodes with the vision line.

This situation is easy to identify when the code is forced to fit the eccentricity
in a circular orbit, since the uncertainties in the fitted values of
$T_p$ and $\omega$ are quite large, and the fitted $e$ value is near 0 with a large uncertainty.
In this case, the best procedure is to re-fit fixing $e=0$.

\subsection{Speed tests and dependence with the number of observations}
\label{subsec:speed}


A primordial motivation for all fitting codes is the execution speed. Some authors claim that SA
is slow. \cite{milone2008} published an extended analysis of classical SA and its performance
using an algorithm similar to \cite{corana1987}, and applied it to the computation of eclipsing
binary stars light curves. But they only fitted light-curves, not RV curves, and used the WD code
to compute the cost function. Fitting three or four parameters, their measured execution times are
as long as $6.2\times 10^4$ s, though in one fit they get a more reasonable value of 386 s.
\cite{prsa2005} claim that SA methods 'are notoriously slow', without going into details.
In their conclusions they state that ASA, with Powell's direction method, is a
'very promising candidate'
\cite{prsa2011} claims that ASA merits applied to eclipsing binaries 'are still to be confirmed'.
\cite{kallrath2006}, in section 4.3.4.3, states for SA that 'the computational cost to reach
this state can be high, as temperature annealing has to be sufficiently slow'.
\cite{wichmann1999} includes SA in {\it Nightfall}, and in their {\it User Manual} states that
'Be prepared for a computing time on the order of a day or more'. We presume that
he implemented a standard SA though no details are given.

To our knowledge, these claims were made in the context of fitting simultaneous photometric
and RV observations of eclipsing binaries where a complex model need to be fitted to account
for the variety of phenomena in such systems. This triggers the number of
parameters to be fitted simultaneously, including tidal distorsions, limb darkening,
gravity brightening, mutual heating effects, computation of visible photosphere due to eclipses,
and stellar spots. It is clear that in these models the cost function is harder to compute and the
space parameter is harder to search. In addition, some of these codes implement classical SA, not
the newer versions with better convergence properties. There have been, however, developments
in computational technology in recent years, which allow computations with much higher speeds.

The computation time of the ASA algorithm depends on a number of factors:

\begin{itemize}

  \item the number of parameters to fit,

  \item the complexity of the functional dependence of the cost function with the parameters,

  \item the size of the dataset, and

  \item the particular configuration of internal variables to run the code, particularly
  $c$ and $N_{gen}$.
\end{itemize}
  In addition, the ASA is a heuristic algorithm with a strong random behaviour.
  This means that each time it is run the trajectory in the parameter space will be different,
  in spite of starting the algorithm with the same initial parameters.

Because of those factors, the precise execution time of the ASA algorithm cannot
be measured or predicted since each run of the code will lead to different values even with
the same dataset and starting points. Thus, we decided to estimate the execution time of
the algorithm taking a statistical approach. For a model radial velocity dataset, we ran
the code a large number of times to obtain statistics on the times needed to complete the fit.
These speed tests were done to provide a complete physical
solution, not only the solution of the ASA algorithm. This includes the code needed
to read the data files, to compute the initial parameter values and, after arriving
to a solution, to compute the physical parameters of the system with their uncertainties,
and to display the plots with the fitted radial velocity curve, the observed data
and the residuals. The tests do not include the computation of the uncertainties using
the MCMC code, which is in separated routines.
The speed tests were done in a DELL inspiron laptop with an Intel Core I5 processor
and 8 GB of RAM memory, running Ubuntu Linux 12.04 LTS.

To assess the dependence of the execution time with the number of RV observations we simulated
diferent datasets with a particular orbital configuration. We chose an orbit with the parameters
$P=10$ days, $T_{p}=2450000.0$, $e=0.1$, $\omega=90$ degrees, $\gamma=0$ km/s, $K_1=20$ km/s.
The chosen orbit depicts a single line system without loss of generality, since for a
double-lined system the total observed points are distributed between the two stars with
the same computational load for computing the cost function as in a single-lined system with the
same total number of points.

\begin{figure}
\epsscale{.80}
\plotone{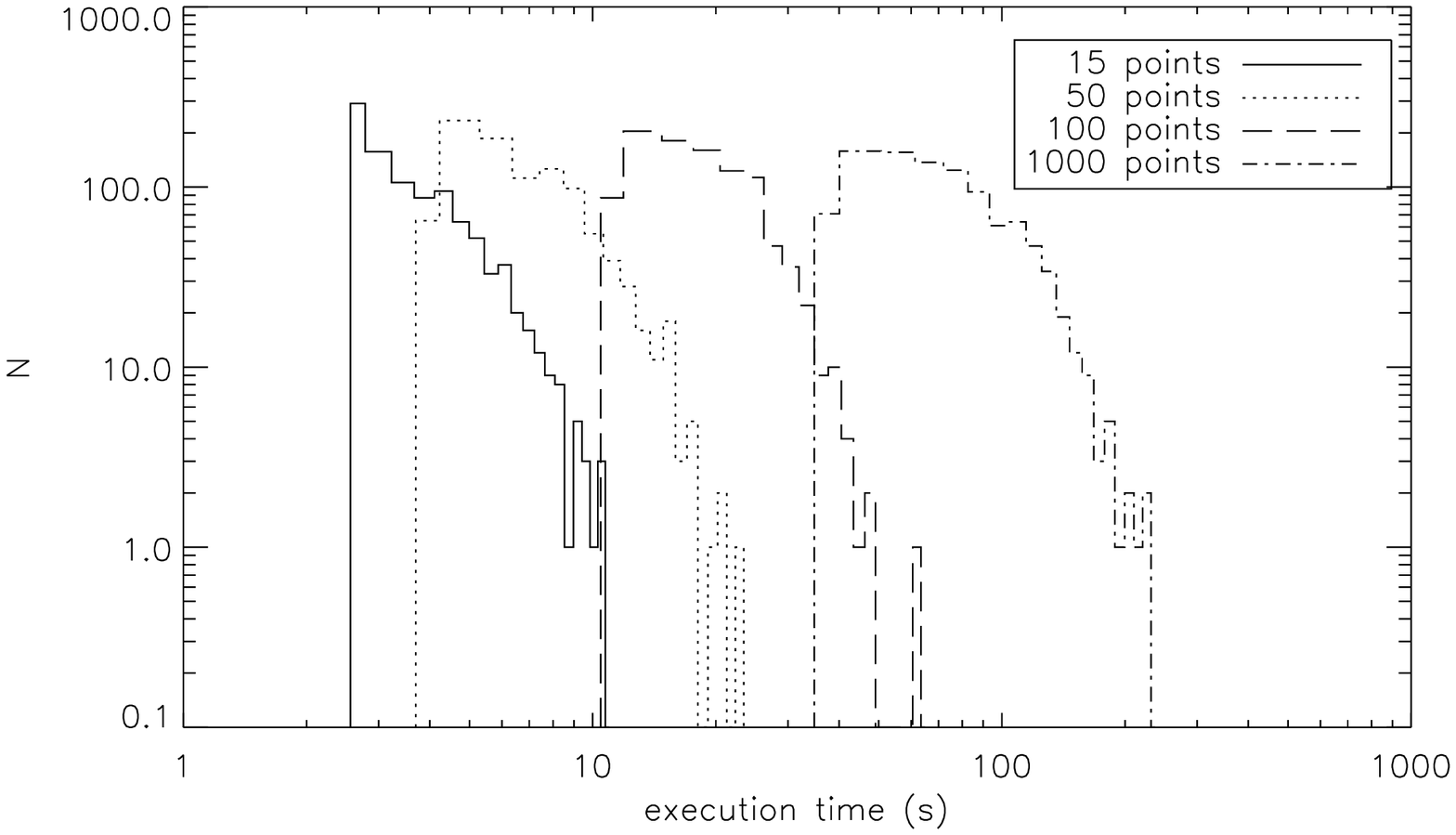}
\caption{Speed test results for SNR=10 synthetic datasets with different number
of points. Each synthetic model was fitted 1000 times to build 20-bin histograms. The abrupt
drop for the lowest bin in each histogram is due to the minimum number of reannealings imposed to
the code for termination and marks the minimum execution time for each dataset.}
\label{fig:execution_times}
\end{figure}

With this configuration we simulated four datasets with 15, 50, 100 and 1000 randomly spaced
data points over a span of 3 periods. Gaussian distributed undertainties with $\sigma_{noise}=2$ km/s
were added to each dataset, resulting in radial velocity curves with a signal-to-noise ratio (SNR) 10.
Each model was fitted 1000 times using \verb;rvfit;. For each run, the execution time and the
$\chi^2$ was logged. Figure \ref{fig:execution_times} shows the execution time histograms of the four
datasets. To quantify this plot we defined $t_{68}$, $t_{95}$, and
$t_{99}$ as the execution times encompassing the 68\%, 95\% and 99\% of all runs.
The results are summarized in Table \ref{tab:execution_times}. Typical execution times are
below 20 s for common single-lined binary datasets and below 30 s for double-line binary
datasets. Earth-like exoplanets with datasets of hundreds of points could be analized in less
than two or three minutes.

\begin{table}[ht]
\begin{center}
\begin{tabular}{cccc}
\hline
\hline
Number of points&\multicolumn{3}{c}{Execution times (s)}\\
		&$t_{68}$&$t_{95}$&$t_{99}$\\
\hline
\hline
15		&4.1	&6.8	&8.9\\
50		&7.6	&12.7	&16.1\\
100		&20.2	&30.4	&38.5\\
1000		&81.1	&132.7	&175.1\\
\hline
\hline
\end{tabular}
\caption{\footnotesize Execution times measured over 1000 consecutive runs for the same
synthetic datasets in Figure \ref{fig:execution_times}. Each time was measured from
the histogram computing the cumulative distribution function (CDF) and interpolating
the desired percentage. Note that doubling the $t_{68}$ execution times is a good guess for
$t_{99}$.}
\label{tab:execution_times}
\end{center}
\end{table}

\subsection{Robustness tests}

To make the code as robust as possible, we performed thousands of fitting tests by
generating synthetic datasets with a wide range of orbital parameters varying
$P$, $Tp$, $\gamma$, $K_1$, and $K_2$ in a aleatory fashion and covering the full range
of values for $e$ and $\omega$. A number of flaws were detected, mainly
related with overflows and zero denominators, which were corrected checking the
conditions of the computations in volved.

\subsection{Refinement of the result}
\label{subsec:refinement}

In their work about the ASA algorithm applied to combined visual and RV observations,
\cite{pourbaix1998} claims that a local search has to be used to tune the minimum found.
This could be due to the fact that the termination condition for his ASA algorithm stops
the computation after a fixed number of temperature reductions with the goal in mind
of obtain a automatic value for the annealing-rate parameter $c$.
Our criterion to stop the annealing loop is somewhat different since
we keep a track of the variations in the cost function $\chi^2$ to have control over
the convergence. So the refinement is an unnecesary step for our algorithm.
Whether the result is to be refined, the output parameters from the ASA algorithm
have to be the input parameters for a code such as Levemberg-Marquadt
\citep[see ][]{wright2009}.

\section{Conclusions}
\label{sec:conclusions}

The Adaptive Simulated Annealing Algorithm (ASA) provide a valuable tool to fit functions in highly-
dimensional parameter spaces. The first versions of the Simulated Annealing algorithm, the so
called Boltzmann Annealing, was computationally slow. The new developments in this algorithm,
namely Adaptive Simulated Annealing (ASA), makes it an option to take into account.

With present domestic computational technologies, a complete solution
for radial-velocity curves with preliminary uncertainties can be obtained in times of the
order of tens of seconds or less. In eclipsing binary systems and transiting exoplanets,
where the period can be determined from their light curves, the fitting is even faster
obtaining a complete solution in seconds.

ASA allows to fit radial velocity curves leaving free all the parameters, including the period.
This is an advantage over others algorithms where the period must be fixed beforehand using other
techniques such as periodograms, Fast Fourier Transforms or Phase Dispersion Minimization techniques.

One of the advantages of the ASA approach is that no derivative is needed to compute the
fit since it is based only on function evaluations. This efficiently avoids local minima.
Also, this approach allows to concentrate all the
physics in one function, the objective function, in our case the $\chi^2$, where the physical
model comes into the $v_{calc}$. More elaborated models can easily be intoduced changing this
function.

Due to these refinements, the advantages of ASA over the classical SA algorithm are clearly stated:
1) Better convergence properties in high dimensionality spaces with great topological complexity,
2) Individual adaptation of each parameter to the local topology, and
3) Faster schedule for the generating and acceptance temperatures.

But this advantages have some cost. The implementation of the ASA algorithm is more
complex than the classical SA algorithm, there is a larger number of internal parameters,
namely, the temperatures and the sensibilities and, 3) the exponential form of the temperature
schedules force to check the exponents to avoid overflows and force the use of double
precission aritmethic.

We have developed a fitting code called \verb;rvfit; which makes use of ASA, to fit keplerian
radial velocity curves. Our code, implemented in IDL, computes initial uncertainties using
a Fisher matrix but we have also tested a routine to obtain uncertainties using a Markov Chain
Monte-Carlo (MCMC) technique with the Metropolis-Hastings sampler.

The \verb;rvfit; code, which is publically avaliable at
\url{http://www.cefca.es/people/~riglesias/index.html},
shows their full capabilities when a search in the full parameter space is needed.
Thus, this code may be helpful in situations where the orbital period is unknown, as
may be the case of RV surveys, since it avoids the previous computation of a periodogram or
the use of other routine to find the period. Also, it provides the full set of parameters in a
consistent way taking into account the observational uncertainties of the data.

Other, simpler techniques, can be applied in the cases where only a small subset of parameters
need to be fitted. For instance, for a circular orbit of known
period, a sine curve can be fitted directly to the radial velocities. This is because there is a
limit to the minimum number of computations that must be done at each temperature to reach the
termination condition. This mechanism is of low efficiency for a small parameters set,
where other algorithms can perform better.

Our tests with real and synthetic radial velocity curves are very promising, with our computed
values and uncertainties in full agreement with the published values, thus proving the
power of this technique.

We note that the current version of \verb;rvfit; is only suited for fitting single exoplanet
systems. Simultaneous multi-planet fits will be implemented in future versions of the code. 
Future refinements of \verb;rvfit; will also introduce Keplerian perturbations in
the orbit due to the presence of other bodies in the system, such as interacting planets
or a thirth star in an external orbit, calculations of jitter contributions for
exoplanet RV datasets and the capability to merge and simultaneously fit RV datasets
from different sources.

\section{Acknowledgements}
This research has been supported by the Spanish Spanish Secretary of State for
R\&D\&i (MICINN) under the grant AYA2012-39346-C02-02. We thank the anonymous referee
for a number of useful comments and suggestions.
This research has made use of the SIMBAD database, operated at CDS, Strasbourg, France,
and of NASA's Astrophysics Data System Bibliographic Services.
We are grateful to Guillermo Torres for fast reply on the data about LV Her.

\clearpage


\begin{thebibliography}{}

\bibitem[Andrae(2010)]{andrae2010} Andrae, R., 2010, arXiv:1009.2755v3

\bibitem[Beavers \& Salzer(1985)]{beavers1985} Beavers, W.I., Salzer, J.J., 1985, \pasp, 97, 355

\bibitem[\v{C}ern\'y(1999)]{cerny1985} \v{C}ern\'y, V., 1985, Optimization Theory and Applications, 45, 41

\bibitem[Chen \& Luk(1999)]{chen1999} Chen, S., Luk, B.L., 1999, Signal Processing, 79, 117

\bibitem[Cochran et al.(2004)]{cochran2004} Cochran, W. D., et al., 2004, \apj, 401, 1029

\bibitem[Coe(2009)]{coe2009} Coe, D., 2009, arXiv:0906.4123v1

\bibitem[Corana et al.(1987)]{corana1987} Corana, A., Marchesi, M., Martini, C., Ridella, S.,
  1987, ACM Trans. Mathematical Software, 13, N3, 262
  

\bibitem[Dreo et al.(2006)]{dreo2006} Dr\'eo, J., P\'etrowski, A., Siarry, P., Taillard, E.,
  \emph{Metaheuristics for Hard Optimization},  2006, p44

\bibitem[Driscoll(2006)]{driscoll2006} Driscoll, P., 2006, Master Thesis, San Francisco State University

\bibitem[Duffett-Smith(1988)]{duffett-smith1988} Duffett-Smith, P., 1998,
  \emph{Practical Astronomy with your calculator, 3nd Ed}, Cambridge University Press

\bibitem[Eastman et al.(2013)]{eastman2013} Eastman, J., Gaudi, B.S., Agol, E., 2005, \pasp, 125, 83

\bibitem[Ford(2005)]{ford2005} Ford, E. B., 2005, \aj, 129, 1706
\bibitem[Ford(2006)]{ford2006} Ford, E. B., 2006, \apj, 642, 505

\bibitem[Gelman et al.(2003)]{gelman2003} Gelman, A., Carlin, J.B., Stern, H.S., Rubin, D.B.,
  \emph{Bayesian Data Analysis, 2nd Ed}, Chapman \& Hall/CRC.

\bibitem[Gim\'enez (2006)]{gimenez2006} Gim\'enez, A., 2006, \apj, 650, 408

\bibitem[Howard et al.(2013)]{howard2013} Howard, A., et al., 2013, Nature, 503, 381

\bibitem[Ingber(1989)]{ingber1989} Ingber, L., \emph{Very fast simulated re-annealing},
  1989, Mathl. Comput. Modelling, 12, 967
  
\bibitem[Ingber(1993)]{ingber1993} Ingber, L., \emph{Simulated annealing: Practice versus theory},
  1993, Mathl. Comput. Modelling, 18, N11, 29

\bibitem[Ingber(1996)]{ingber1996} Ingber, L., \emph{Adaptive simulated annealing (ASA): Lessons learned},
  1996, Control and Cybernetics, 25, N1, 33
  
\bibitem[Kallrath \& Milone(2006)]{kallrath2006} Kallrath, J., Milone, E. F., 2006,
  \emph{Eclipsing Binary Stars.  Modelling and Analysis}, Springer

\bibitem[Kirkpatrick et al.(1983)]{kirkpatrick1983} Kirkpatrick S., Gelatt Jr. C., Vecchi M.,
  1983, Sci, 220 (4598), 671

\bibitem[K\"urster et al.(2008)]{kurster2008} K\"urster, M., Endl, M., Reffert, S., 2008, A\&A, 483, 869
  
\bibitem[Levenberg(1944)]{levenberg1944} Levenberg, K., 1944, Quarterly of Applied Mathematics, 2, 164

\bibitem[Locatelli(2002)]{locatelli2002} Locatelli, M., \emph{Simulated annealing algorithms for
  continuous global optimization}, Handbook of Global Optimization II, Kluwer Academic Publishers, p179

\bibitem[L\'opez-Morales \& Ribas(2005)]{lopez-morales2005} L\'opez-Morales, M., Ribas, I., 2005, \apj, 631, 1120

\bibitem[MacKay(2006)]{mackay2006} MacKay, D.J.C., 2006, \emph{Information Theory,
    Inference, and Learning Algorithms}, Cambridge University Press

\bibitem[Marquardt(1963)]{marquardt1963} Marquardt, D., 1963, SIAM Journal on Applied Mathematics, 11 (2), 431

\bibitem[Meeus(1998)]{meeus1998} Meeus, J., 1998, \emph{Astronomical Algorithms, 2nd Ed}, Willmann-Bell Inc.

\bibitem[Meschiari \& Laughlin(2010))]{meschiari2010} Meschiari, S., Laughlin, G.P., 2010, \apj, 718, 543

\bibitem[Meschiari et al.(2009))]{meschiari2009} Meschiari, S., Wolf, A.S., Rivera, E., Laughlin, G.P.,
  Vogt, S., Butler, P., 2009, \pasp, 883, 1016

\bibitem[Metropolis et al.(1953)]{metropolis1953} Metropolis N., Rosenbluth A., Rosenbluth M.,
Teller A., Teller E., 1953, J. Chem. Phys. 21(6), 1087

\bibitem[Milone \& Kallrath(2008)]{milone2008}  Milone, E.F, Kallrath, J., 2008,
  \emph{Short-Period  Binary Stars. Observations,Analyses and Results}, 
  Milone, E. F., Leahy, D.A., Hobill, D.W., Springer, p191

\bibitem[Murty(1983)]{murty1983} Murty, K.G., 1983,
  \emph{Linear programming}, New York: John Wiley \& Sons Inc


\bibitem[Nelder \& Mead(1965)]{neldermead1965} Nelder, J.A, Mead, R., 1965, Computer Journal 7, 308

\bibitem[Ohta et al.(2005)]{ohta2005} Ohta, Y., Taruya, A., Suto, Y., 2005, \apj, 622, 1118

\bibitem[Otten \& van Ginneken(1989)]{otten1989} Otten R.H.J.M., van Ginneken L.P.P.P., 1989,
  \emph{The Annealing Algorithm}, Kluwer Academic Publishers
  
\bibitem[Pepe et al.(2013)]{pepe2013} Pepe, F., et al., 2013, Nature, 503, 377

\bibitem[Pincus(1970)]{pincus1970} Pincus, M., 1970, Operations Research, 18, 1225

\bibitem[Pourbaix(1998)]{pourbaix1998} Pourbaix, D., 1998, A\&AS, 131, 377

\bibitem[Press et al.(1992)]{press1992} Press W., Teukolsky S., Vetterling W., Flannery B., 1992,
  \emph{Numerical Recipes in C, 2nd ed}, Cambridge University Press

\bibitem[Pr\v{s}a (2011)]{prsa2011} Pr\v{s}a, A., 2011,
  \emph{PHOEBE Scientific Reference}, Philadelphia: SIAM

\bibitem[Pr\v{s}a \& Zwitter(2005)]{prsa2005} Pr\v{s}a, A., Zwitter, T., 2005, \apj, 628, 426

\bibitem[Salamon et al.(2002)]{salamon2002} Salamon, P., Sibani, P., Frost, R., 2002,
  \emph{Facts, Conjectures, and Improvements for Simulated Annealing}, Philadelphia: SIAM

\bibitem[Sanchis-Ojeda et al.(2013)]{sanchis-ojeda2013} Sanchis-Ojeda, R., et al., 2013, \apj, 774, 54

\bibitem[Sinnott(1985)]{sinnott1985} Sinnott, R.W., 1985, Sky and Telescope, 70, 159

\bibitem[Sybilski et al.(2013)]{sybilski2013} Sybilski, P., Konacki, M., Koz0'0owski, S. K.,
  Helminiak, K. G., 2013, \mnras, 431, 2024

\bibitem[Szu and Hartley(1987)]{szu1987} Szu, H., Hartley, R., 1987, Phys. Lett. A, 122, 3-4, 157

\bibitem[Torres et al.(2009)]{torres2009} Torres, G., Sandberg Lacy, C. H., Claret, A., 2009, \aj, 138, 1622

\bibitem[Wichmann(1999)]{wichmann1999} Wichmann, R. 1999,
  \emph{Nightfall Users Manual}

\bibitem[Wichmann(2011)]{wichmann2011} Wichmann, R. 2011,
  \emph{Nightfall: Animated Views of Eclipsing Binary Stars}, Astrophysics Source Code Library, ascl:1106.016

\bibitem[Wilson \& Devinney(1971)]{wilson1971} Wilson, R. E., Devinney, E. J., 1971, \apj, 166, 605

\bibitem[Wright \& Howard(2009)]{wright2009} Wright, J. T., \& Howard, A. W., 2009, \apjs, 182, 205

\end{thebibliography}
\end{document}